\documentclass[fleqn,usenatbib]{mnras}

\usepackage{graphicx}	
\usepackage{amsmath}	
\usepackage{nccmath}
\usepackage{amssymb}	
\usepackage{caption}
\usepackage{subcaption}

\usepackage{smartdiagram}
\usesmartdiagramlibrary{additions}
\usepackage{multirow}

\newcommand{\Epar}{$E_{\parallel } $}

\title[Synchro-curvature emitting regions]
{Synchro-curvature emitting regions in high-energy pulsar models}

\author[\'I\~niguez-Pascual, Vigan\`o, \& Torres]{
Daniel \'I\~niguez-Pascual$^{1,2}$\thanks{E-mail: iniguez@ice.csic.es}, Daniele Vigan\`o$^{1,2}$\thanks{E-mail: vigano@ice.csic.es},
Diego F. Torres$^{1,2,3}$\thanks{E-mail: dtorres@ice.csic.es}
\\
$^{1}$Institute of Space Sciences (ICE, CSIC), Campus UAB, Carrer de Can Magrans s/n, 08193 Barcelona, Spain\\
$^{2}$Institut d’Estudis Espacials de Catalunya (IEEC), 08034 Barcelona, Spain\\
$^{3}$Institució Catalana de Recerca i Estudis Avançats (ICREA), E-08010 Barcelona, Spain \\
}

\date{}

\pubyear{2022}

\begin{document}
\label{firstpage}
\pagerange{\pageref{firstpage}--\pageref{lastpage}}
\maketitle

\begin{abstract}

The detected high-energy pulsars' population is growing in number, and thus, having agile and physically relevant codes to analyze it consistently is important. 
Here, we update our existing synchro-curvature radiation model by including a better treatment of
the particle injection, particularly where the large pitch angle particles dominate the spectra, and by implementing a fast and accurate minimization technique. 
The latter allows a large improvement in computational cost, needed to test model enhancements and to apply the model to a larger pulsar population.
We successfully fit the sample of pulsars with X-ray and $\gamma$-ray data.
Our results indicate that, for every emitting particle, the spatial extent of their trajectory where the pitch angle is large and most of the detected X-ray radiation is produced is a small fraction of the light cylinder. 
We also confirm with this new approach that synchrotron radiation is not negligible for most of the gamma-ray pulsars detected.
In addition, with the results obtained, we argue that J0357+3205 and J2055+2539 are MeV-pulsar candidates and are suggested for exhaustive observations in this energy band.
\end{abstract}


\section{Introduction}\label{introduction}

Pulsars are fast-spinning neutron stars, usually detected in radio when their radiation beam crosses our line of sight. The current population of radio pulsars is over 3300 \citep{ATNF-Catalog}.
Within this sample, there are about 300 
pulsars detected in $\gamma$-rays and about a few dozen 
detected in X-rays. 
The increase in the number of high-energy pulsars discovered along the last decade is mostly attributed to the \emph{Fermi} mission, which has discovered and characterized pulsars in the $\sim 0.1-100$ GeV range.
The latest Fermi-LAT pulsar catalog \citep{2fpc} is expected to be updated soon.

Despite having been detected for over 50 years, many aspects of pulsar physics are still unclear. Among them, here we focus on the origin of their high-energy emission.
This radiation is produced in the magnetosphere of pulsars by field acceleration of charged particles up to high energies (see  \cite{pulsar_electrodynamics} for initial ideas). Such acceleration is possible only in specific places of the magnetosphere, where the force-free condition locally breaks down.
What is their location and how large are such accelerating regions? Classical models (see e.g. the pioneering works by \cite{arons83,Cheng86}) have assumed them close to the separatrix (i.e., the field line between the open and closed magnetic field lines of a rotating dipole). 
However, more recent results from particle-in-cell (PIC) simulations point to the Y-point, a localized region beyond the light cylinder where the equatorial magnetic field lines reconnect (\cite{Cerutti19} and references in it).
Complementary to the location, radiative models aim at reproducing the spectral energy distributions (SED) of the observed high-energy radiation. 
They require specifying the plasma particles' kinetic properties, which determine the emission according to classical electrodynamics.
Confronting these models with the spectra of $\gamma$-ray pulsars links  phenomenology with the underlying physical mechanism.

In this paper we update and improve over an existing synchro-curvature radiation model \citep{outer_gap_model_paper_2,diego_solo}.
Such model relies on an effective parametrization of the relevant physical quantities that affect the particle dynamics.
The synchro-curvature model has first been used to successfully fit the $\gamma$-ray data of all $\gamma$-ray pulsars  \citep{Vigan_2015, Vigan_2015b}. 
It was lately extended to include also the X-ray band.
A single and austere model (just three main physical parameters and a normalization) was shown to be able to describe the X-ray and $\gamma$-ray spectra of pulsars reasonably well \citep{diego_solo}.
The model explained the appearance of sub-exponential cutoffs at high energies as a natural consequence of synchro-curvature-dominated losses, and  the flattening of the X-ray spectra at soft energies as a result of propagating particles being subject to synchrotron losses all along their trajectories. 
Using the model to describe both the X-ray and gamma-ray emission for all detected pulsars showed that synchrotron emission plays a significant role in the spectral formation \citep{diego_solo,systematic_2019}
As spinoff, the model can be used to predict the level of X-ray emission of $\gamma$-ray pulsars without a clear counterpart at this wavelength, and has lead to new detections \citep{Li18}.

The rest of the paper is organized as follows. 
Section \ref{sc_model}
describes details of the model, including an improved implementation for computing the averaged spectra, specially around any particles' injection point.
Section \ref{fitting_approach} describes a new minimization algorithm used to find the best-fit parameters of a pulsar. 
Section \ref{application} presents a systematic fitting of the known high-energy pulsars' spectral energy distribution (SED), as well as a discussion of the accelerating and the injection regions per se.
The size and location of the injection region is discussed in Section 
\ref{acc-reg}.
Finally, 
Section \ref{conclusions} briefly draws our main conclusions.

\section{Synchro-curvature radiation model with extended injection region}\label{sc_model}

The model uses the synchro-curvature radiation formalism developed in
\citep{compact_formulae}, and stems from our previous works \citep{outer_gap_model_paper_1, outer_gap_model_paper_2, Vigan_2015, Vigan_2015b, diego_solo, systematic_2019}, to which we refer the reader for further details.
%

\subsection{Free parameters}

The minimal model is based on the dynamics of an ensemble of particles traveling along a magnetic field line in an acceleration region of the magnetosphere.
The calculation is done in terms of three free parameters: \\
(i) the electric field parallel to the magnetic field, $E_{\parallel}$, assumed constant throughout an accelerating region; \\
(ii) the local magnetic field $B$, which is assumed as a function of the pulsars' measured spin period, $P$, and its time derivative, $\dot{P}$, the distance $x$ along the field line, and the magnetic gradient $b$ assumed as a free parameter: $B=B_s(x/R_s)^{-b}$, where $B_s=6.4\times 10^{19} \sqrt{P \dot{P}}$ is the often-used timing-inferred estimate of the surface polar magnetic field (assuming a perfect dipolar configuration of the magnetic field, in which the field decreases with distance as $B \propto x^{-3}$), and $R_s$ is the radius of the neutron star;\\
(iii) and the spatial extent of the observationally-relevant emitting region for particles injected in $x_{in}$, $x_0/R_{lc}$, which enters by means of a weighting function representing the  reduction of the number of emitting-particles directed towards the observer at a distance from their injection point. \\

Moreover, we assume an effective simple parametrization of its curvature radius, $r_c=(x/R_{lc})^{\eta}$. In  previous works we have shown that the value of $\eta$ have a negligible impact on the spectral prediction compared to other parameters. Therefore, here as well as in previous works, we have fixed $\eta=0.5$ without loss of generality.

In this paper we consider an additional parameter, $x_{lp.in}$, related to the injection region, as explained below.

\subsection{Equations of motion}

For given values of $P$, $\dot{P}$, $E_\parallel$, $\eta$, $B_s$, and an assumed initial pitch angle, the model solves the equation of motion of the particles to gather their trajectories: 
\begin{equation}
 \frac{d\vec{p}}{dt} = ZeE_\parallel \hat{b} - \frac{P_{sc}}{v}~\hat{p}~.
\end{equation}
Here we can decompose the momentum $\vec{p}$ into its parallel ($p_\parallel=p\cos\alpha$) and perpendicular components ($p_\perp = p\sin\alpha$) respect to the magnetic field lines (directed along $\hat{b}$); $Ze$ is the charge of the particle, $v\sim c$ its velocity and $P_{sc}$ is the synchro-curvature power (see Appendix \ref{app:model} for the $dP_{sc}/dE$ expression). From these equations, one can determine the evolution of the Lorentz factor $\Gamma$, the synchro-curvature parameter $\xi$, and the pitch angle $\alpha$ (see e.g., \cite{compact_formulae} for details).

\subsection{Pitch angle of particles}

In our approach, the pitch angle evolves following the equations of motion. In general, the pitch angle steeply declines since the perpendicular momentum is efficiently radiated away by synchrotron emission. Quantitative details on the evolution of $\sin{\alpha}$ along the trajectory were shown before in e.g.,  the middle panel of Fig. 2 of \cite{compact_formulae}.

In general, we assume for simplicity that all particles have the same initial pitch angle, $45^o$, when they start to be accelerated. A more realistic approach would be to assume that each particle has a different pitch angle, of course. However, due to the fast decrease of the pitch angle, the precise value of its initial value is pretty much irrelevant for the fit.
As an example, in the case of Geminga, changing the values of $\alpha_{in}$ introduces minimal differences among the best-fit parameters of $E_\parallel$ and $x_0$, and differences regard basically the value of the local magnetic field, through $b$ (going e.g., from a pitch angle of 45$^0$ to 1$^0$ (45$^o$ to 99$^0$), there is a 30\% (2\%) change in the value of $b$).
This behavior is general. Differences in the best-fit local magnetic field are only appreciable when $\alpha_{in}$ is very small. Since this is not expected to happen for all particles, introducing further complications in our effective model to deal with an initial pitch angle distribution is not warranted.
Indeed, we do not assume that particles enter the acceleration region with a direction along the magnetic field line. On the contrary, since the local magnetic field far from the surface (see Table 1 below) is much weaker than the critical quantum magnetic field, pairs are arguably produced mostly by photon-photon interactions, so that the initial angle would expected to acquire random values, ${\cal O}(1)$.

Note that, for the sake of simplicity, we omit more complex effects that could affect the pitch angle evolution, like resonant absorption of radio photons. This could provide the traveling charged particles with a non-negligible pitch angle along the trajectory \citep{Lyubarskii1998,Harding2008}, beyond the initial injection. In our case, any possible pitch angle dependence is only effectively modeled by  the averaging of the particles spectrum  
in which large pitch angle particles are important (see below).

\subsection{Particle distribution}

At any position along the trajectory, the emission of radiation  ${dP_{sc}}/{dE}(B,E_\parallel,r_c,\Gamma,\alpha)$ is completely specified by: (i) the local kinetic properties of the particles (Lorentz factor $\Gamma$, pitch angle $\alpha$, position), and (ii) the local properties of the magnetic field (intensity and curvature radius).
See Appendix \ref{app:model} 
for more explicit expressions.
Note that the trajectory and the consequent spectrum are supposed to be effective averages over the diversity of magnetic field lines and trajectories within the acceleration region, including possibly the emission of particles going backwards. Our absence of a detailed geometrical model only allows us to model this in an effective way:
The contributions at a given position of the trajectory are weighted by an effective particle distribution function $dN/dx$, that represents particles whose radiation is directed towards the observer. As in our previous works, we adopt:
  \begin{equation}
  \frac{dN_e}{dx}= \left [ \frac{N_0}{x_0 (1 - e^{-(x_{out}-x_{in})/x_0}) } \right]
  e^{-(x-x_{in})/x_0}.
  \label{eq:effective_particle_distribution}
\end{equation}
Here, 
$N_0$ is the {\it normalization}, such that
$\int_{x_{in}}^{x_{out}} (dN_e/dx) dx=N_0; $
and
$x_0$ is a length scale. $(x_{out}-x_{in})$ is the extent where the parallel accelerating field is assumed to exist.
The ratio $x_0/(x_{out}-x_{in})$, or its inverse, called contrast, is a measure of how uniform is the distribution of particles injected in a given $x_{in}$ that are still emitting towards us at a distance $x$ along the magnetic field line. 
The larger $x_0$ is, the more uniform is this distribution. 
Note that the $N_0$ value does not affect the shape of the spectrum but only its absolute scaling.  For each set of values of the model parameters ($E_{\parallel}$, $x_0$,  $b$ and $x_{lp.in}$), $N_0$ is simply found by linear regression as explained below.

We consider that 
$x_{out}-x_{in} = x_{\rm max} = R_{lc}$.
Note however that the precise value of $x_{\rm max}$ is irrelevant for the fit presented next, as long as $x_{max}\gg x_0$.
To simplify,
$x_{in} $ and $x_{out} $  are assumed the same for all pulsars. 
We consider accelerating region sizes equal to $R_{lc}$ placed around the light cylinder, leaving the contrast free, in order to constrain the scale of the emission region that is relevant for us as observers once particles are injected.

We have also earlier investigated in our papers that the influence of $x_{in}$ is negligible  when it is chosen in a broad range, and this is why we first fix it for all pulsars to the same value, $x_{in} = 0.5 R_{lc}$, to simplify. See below for a more explicit discussion on this point.

The total spectrum is thus defined by
    \begin{equation}
        \frac{dP_{tot}}{dE} = \int_{x_{in}}^{x_{out}} \frac{dP_{sc}}{dE} \frac{dN}{dx} dx
        \label{eq:convolved_power_spectra}
    \end{equation}
Since the effective parameters are thought to be a-priori agnostic, the fits to real data can constrain the physical assumptions.

\subsection{Particles' birthplace treatment}
\label{birth-treat}

When particles are created, their pitch angle is non-negligible, and thus the synchrotron-like emission is strong. Therefore, the X-ray SED is mostly shaped around at innermost region of the particles' trajectories, whereas the $\gamma$-ray spectrum is less sensitive to the birth properties, since it is formed mostly farther from it, where the electric acceleration provide $\Gamma \sim 10^6-10^8$ and $\alpha \sim 0$ \citep{outer_gap_model_paper_2}.
In previous works, we had adopted  a simplified numerical treatment assuming that, 
while particles were in the innermost part of their trajectory, the contributions to the spectra were dominated by the particles' birth properties (initial pitch angle and $\Gamma$).
The size of this innermost region for a given injection point was estimated from the trajectories of particles in a per-pulsar basis, as the extent at which a particle injected at the start of the accelerating region, and subject to the corresponding losses provided by the local magnetic field and curvature, would have traversed at least $10^{-4} R_{lc}$. 
This seems to be a reasonable consideration given the smallness of the observationally-relevant portion of the accelerating region for a given particle as obtained a-posteriori in the fits (i.e., the comparison between parameter $x_0$ and the relevant spatial scale $R_{lc}$ used to describe the scenario).
We shall refer to this fixed determination of the innermost particle trajectories' properties as the {\it minimal model} below. In order to test the validity of this assumption, we shall here leave the spatial scale of this
region, as an additional free parameter. We shall call this scale $x_{lp.in}$, to represent that it is related to the large pitch angle dominance region for a given injection point.

In both the minimal model with estimated $x_{lp.in}$ and the one in which $x_{lp.in}$ will be a free parameter, particles are injected at trajectory positions in a region close to $x_{in}$ (see Section 5 for additional discussion on the injection region treatment) and travel in the same direction through the acceleration funnel.
At every position of the trajectory there will be particles that have been injected at different (earlier) positions and that have traveled different distances, thus emitting differently.
We shall no longer consider that there is an innermost region (of any size) in which the emission of particles created  at any point overcomes that of particles propagating from earlier birthplaces. 
Instead, we shall compute the spectra by introducing an averaging of the single-particle spectra, in a similar way as what was done with the particle distribution itself above, Eq. \eqref{eq:effective_particle_distribution}.
The averaged single-particle spectra in the position $x$ is to be defined as:
\begin{equation}
    \left\langle\frac{dP_{sc}}{dE}(x)\right\rangle = \frac{1}{x_{lp.in}} \int_0^{min(x, x_{lp.in})} \frac{\Delta\Omega(x_{nb})}{\Delta\Omega(0)}\frac{dP_{sc}}{dE}(x - x_b)dx_b
    \label{eq:new_averaging}
\end{equation}
where
\begin{equation}
\centering
    x_{nb} = min(x, x_{lp.in}) - x_b ,
\end{equation}
and we set:
\begin{equation}
    \frac{\Delta\Omega(x)}{\Delta\Omega(0)}
    = \frac{\sin\alpha(x)}{\sin\alpha(0)}\frac{\Gamma(0)}{\Gamma(x)}~,
\end{equation}
which is a monotonously decreasing function, that goes like $\sim x^{-2}$ for $x\sim 10^{-4}$-$10^{-1} R_{lc}$ \citep{compact_formulae}.

Such weighting factor can be justified by a combination of factors. 
On the one side, it relates to the probability of detecting such radiation, which is emitted over an angular surface $\Delta \Omega(x) \sim 2 \pi \sin(\alpha(x))\Delta \theta(x)$, where $\Delta \theta(x) \sim 1/\Gamma(x)$ (according to classical electrodynamics). 
On the other side, we are considering a unique value of initial $\alpha$ and $\Gamma$ as averages. But since each particle will have their own values, and the emitted spectrum is sensitive to them, as it would also be to the direction (backward or forward) in which they travel, $\Delta \Omega(x)$ gives the model a chance to effectively accommodate variations. 
Once the averaged-spectrum at each position, Eq. \eqref{eq:new_averaging}, is known, it is weighted with the effective particle distribution, Eq. \eqref{eq:effective_particle_distribution}:
    \begin{equation}
        \frac{dP_{tot}}{dE} = \int_{0}^{x_{max}} \left\langle\frac{dP_{sc}}{dE}(x)\right\rangle \frac{dN}{dx}(x)~ dx
        \label{eq:fully_convolved_power_spectra}
    \end{equation}
With this formalism we have changed the way of computing the spectra, from the assumption in which all particles predominantly emitted with their birth spectrum in an inner region of fixed size, to a weighted average over all possible spectra produced by particles injected at different points close to $x_{in}$.
We did so by adding a single extra parameter $x_{lp.in}/R_{lc}$ and an additional weighting function.

Note then that $x_{lp.in}$ represents the spatial scale, around a given $x_{in}$, where the average spectrum is dominated by particles with relatively large pitch angle. The model does not need nor require that the acceleration of particles up to
 $>10^6$ V/cm happens within $x_{lp.in}$. 
The equations of motion --which we self-consistently solve for each pulsar- give the evolution of the Lorentz factor along the accelerating region (see the discussion below for an explicit example of the particle properties).

\section{Minimization algorithm } \label{fitting_approach}

Our earlier works used a gridding over three model parameters (the normalization can be analytically computed as a multiplicative factor) in order to fit the pulsars' SED. The typical fitting time per pulsar was several hours (5 to 10).
Our aim is to fit the increasingly numerous high-energy pulsar population to extract global features, as well as to test other possible physical improvements over the model in the future. %
Thus, in order to keep the computational costs affordable, we have left the gridding and implemented a  Nelder-Mead minimization  \cite{nelder_mead}.

The Nelder-Mead algorithm (also called downhill simplex method, see \cite{numerical_recipes}) only involves the evaluation of the function (not its derivatives) and is based on the geometrical figure of a simplex.
A simplex is the generalization to $N$ dimensions of a triangle and thus has $N + 1$ vertices, being $N$ the dimension of the problem.
Each vertex is a point in an $N$-dimensional parameter's space, with the coordinates being the parameters of the function to minimize.
This function is evaluated at each vertex, promoting a motion of the simplex through the parameter's space with the aim of minimizing the function's return values.
In our problem, the parameters are (\Epar, $x_0/R_{lc}$, $b$), and the function to minimize is the reduced $\chi^2$, $\overline{\chi^2}$, computed on the observed luminosity:
\begin{equation}
    \overline{\chi^2} = \frac{1}{N_{bin} - 3} \sum_{bin} \frac{(L^{bin}_{obs} - L^{bin}_{theo})^2}{(\delta L^{bin}_{obs})^2}
    \label{eq:chi2}
\end{equation}
where $N_{bin}$ is the number of data points, $L_{theo}^{bin}$ is the theoretical luminosity produced by our model,  $L_{obs}^{bin}$ is the observational luminosity and $\delta L_{obs}^{bin}$ is the experimental error of the data. The value of 3 stands for the number of free parameters. Note that we have an additional parameter, $x_{lp.in}$, when we leave free the spatial extent of the region of large pitch angle dominance, as explained in \S \ref{fit_x_lpin}.

The fastest way to a solution begins with implementing a two-dimensional (2D) Nelder-Mead minimization having $\log_{10} (E_{\parallel})$  and $\log_{10}(x_0/R_{lc})$ as free parameters.
Afterwards, we do a one-dimensional (1D) minimization with a bisection method, to search for $b$, having $E_{\parallel}$ and $x_0/R_{lc}$ fixed to the values found in the 2D minimization done before. 
In this way, we end up with an initial set of values of the three model parameters which is likely close to the solution, and with that we do a 3D Nelder-Mead minimization with all free parameters at once to nail it down.

The reason why this approach is better than doing directly a 3D minimization from the start is that $E_{\parallel}$ and $x_0/R_{lc}$ drive the spectral shape in the $\gamma$-ray band, while $b$ mainly affects the X-ray band.
Thus, we can iteratively approach the whole solution, and only when sufficiently close to it, go into the more costly part of the minimization with all parameters onboard. 
In this way, also the risk of falling into local minima is minimized.

Before the algorithm starts, an initial simplex has to be defined.
All its vertices are built from an initial point $\Vec{p_0}$, as follows: $\Vec{p_i} = \Vec{p_0} + \lambda \cdot \Hat{e_i}$, where $\lambda$ is the characteristic length of the problem (we choose it to be $0.1$) and $\Hat{e_i}$ are unit vectors with $i =1, \dots, n$. 
The closer the simplex is initialized with respect to the real minimum, the fewer steps will be needed for the algorithm to find the minimum, i.e. the smaller the computational time will be.
Therefore, we have to find convenient coordinates for the initialization $\Vec{p_0}$ (that is, values of $E_{\parallel}$, $x_0/R_{lc}$ and $b$). 
The existence of some hints about the model parameters could help in this task.
For that we used the fact, as shown by \cite{Vigan_2015b} with gamma-ray data and by \cite{diego_solo} with both X-rays and gamma-day data, that
$E_{\parallel}$ is anti-correlated with $x_0$, with a best-fit given by $\log x_0 \sim -1.28 \log E_{\parallel} + 16.9$ (with non-negligible dispersion).
Similarly, a reasonable correlation exists between $E_{\parallel}$ and the light cylinder radius, defined as $B_{lc}= 5.9 \times 10^8 P^{-2.5}  \Dot{P}^{0.5}$. The best-fit relation was found to be 
$\log E_{\parallel} \sim 0.78  \log_{10}B_{lc} + 4.95$.
The initial value of the parameter $b$ is fixed here to a fiducial value of $3$, qualitatively expected for a dipolar magnetic field and close to the typical best-fit values found in previous studies.

Finally, the normalization $N_0$ allows to scale the theoretical spectrum without affecting its shape in accordance to the observational one. Its best-fit value is analytically computed by linear regression during the fitting procedure, for any given set of the other free parameters:
\begin{equation}
    N_0 = \dfrac{\sum_{bin} \dfrac{L^{bin}_{obs} \cdot L^{bin}_{theo}}{(\delta L^{bin}_{obs})^2}}{\sum_{bin} \dfrac{(L^{bin}_{theo})^2}{(\delta L^{bin}_{obs})^2}}
    \label{eq:normalization_N0}
\end{equation}
which secures minimizing $\chi^2$ for that particular model incarnation.

The algorithm stops when the fractional distance in $\overline{\chi^2}$ between the vertices of the simplex with higher and lower function evaluations is smaller than $10^{-4}$.
This represents a proper balance between precision and computational time.
The errors of the model parameters are estimated following the prescriptions of \cite{avni_factor}, 
which defines $\overline{\chi^2}_{limit}$, a 1 $\sigma$ confidence interval for the corresponding degrees of freedom.

With this new approach, we have been able to reduce the computational time required for the spectral fitting of a high-energy pulsar using three parameters to less than half an hour (with only 2 parameters, $E_{\parallel}$ and $x_0$ it is reduced to just a few minutes).
In addition, the new approach brings an improvement in the accuracy of the best-fit parameters. 
While with a scanning of parameters the accuracy in the determination of the model parameters was sometimes given just by the grid distance (typically, e.g., 0.02 for $\log E_\parallel$), here it can lower.

\section{Application to the high-energy pulsar population}
\label{application}

\subsection{Sample}

We study a population of high-energy pulsars consisting of 36 objects.
30 of these are normal pulsars (PSRs), with $P > 10$ ms, and
6 are millisecond pulsars (MSPs), with $P < 10$ ms. 
Three high-energy pulsars, PSR (J2043+2740) and
two MSPs (J1824-2452A and J1939+2134) are not considered
in the study, due to their low data quality.
We perform a systematic fitting of our synchro-curvature model to the available observational data of this sample of high-energy pulsars.
The dataset used is that of  \cite{CotiZelati20}. We note 
that the (very) high energy tails data of some of the pulsars, e.g., Vela \cite{HESSVela200} or \cite{MAGIC-2020-Geminga}, correspond to the peaks of their respective light curves and are not averaged spectra as the GeV emission and the X-ray emission are. Mixing phase-averaged spectra with some obtained at particular phases is not consistent. However, we notice that in general the high-energy tails have a  high flux level to be fitted by the same process or the same population of particles or the same electric field than what produces the rest of the spectra in our model, and e.g., it may be ascribable to self-synchrotron Compton or inverse Compton mechanisms.

\subsection{Best-fit results resulting from the averaged inner region}

\begin{figure*}
        \centering
        \includegraphics[width=0.33\textwidth]{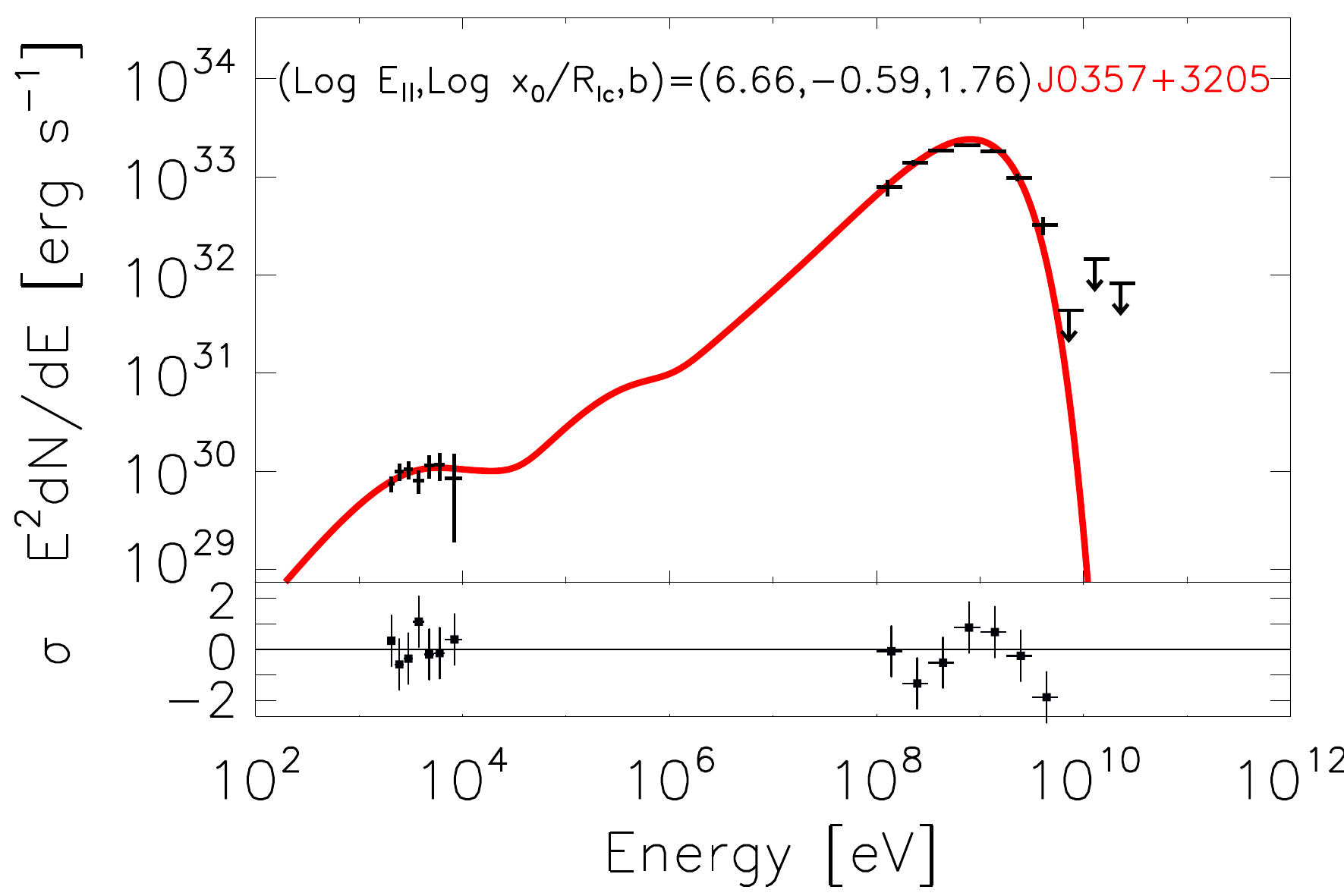}
        \includegraphics[width=0.33\textwidth]{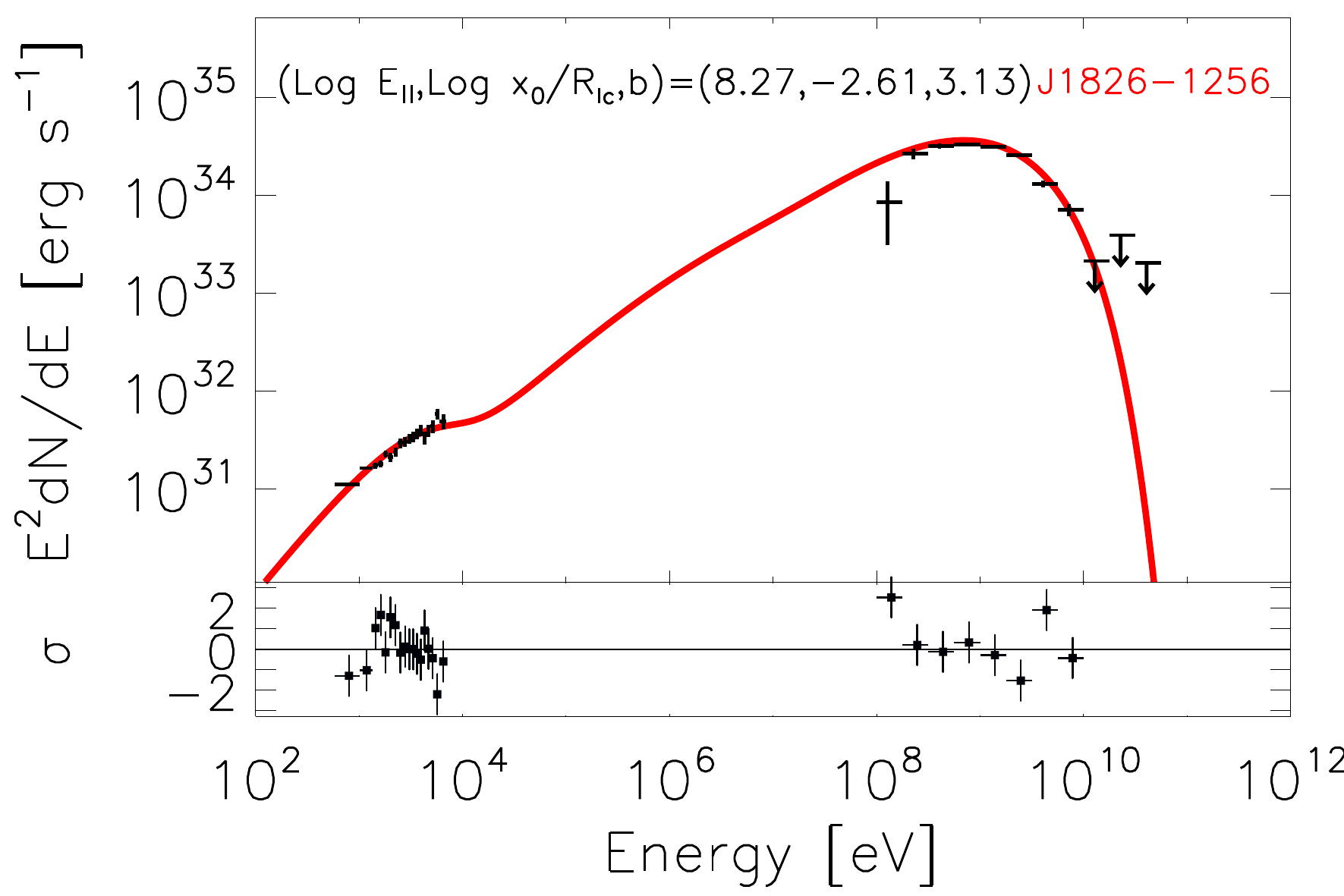}
        \includegraphics[width=0.33\textwidth]{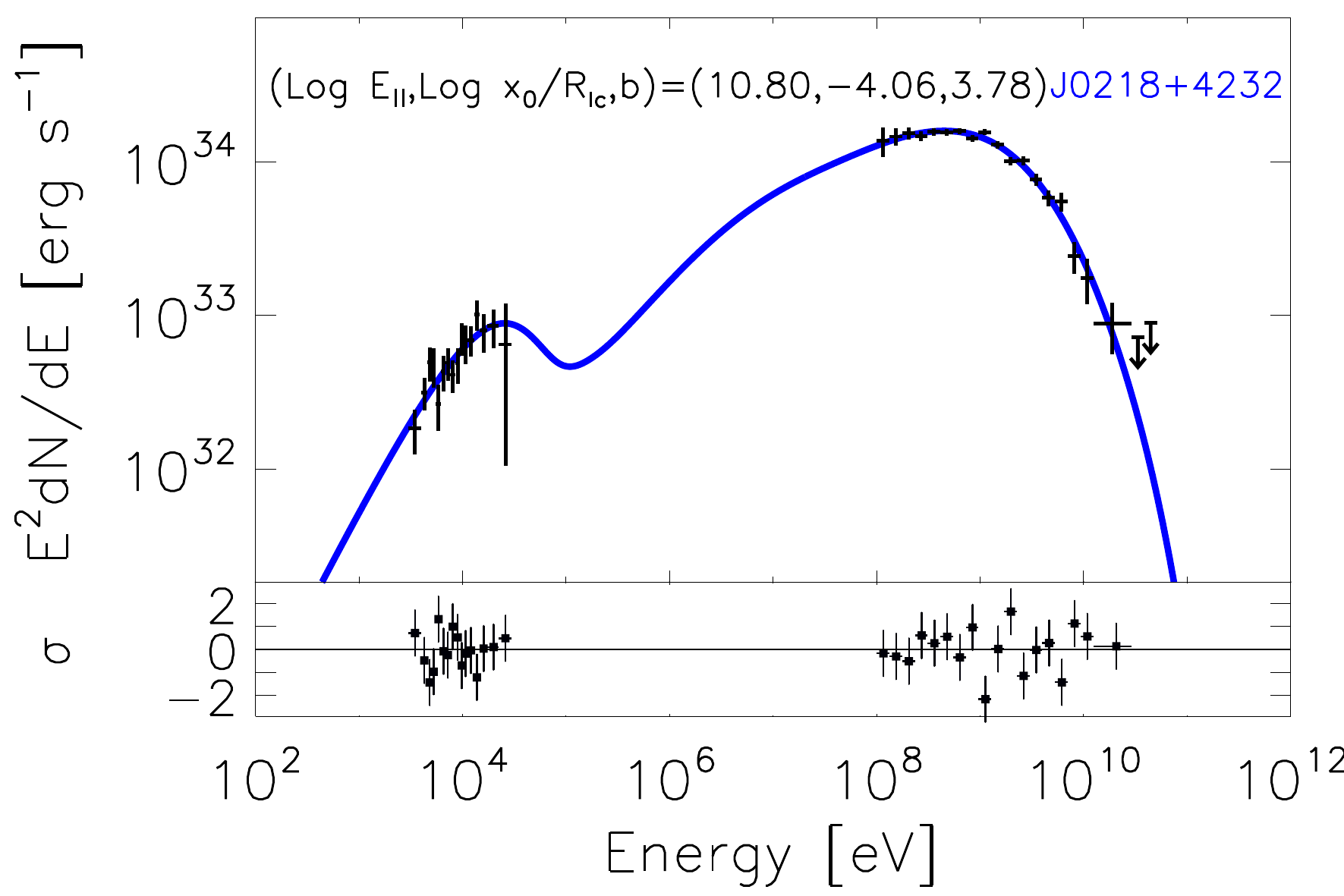}
    \caption{The two left panels show the minimal model SEDs of J0357+3205 and J1826-1256 for which the averaging of the inner spectra now allow for a description with just one set of parameters (\Epar, $x_0/R_{lc}$, $b$). The right panel shows the case of a millisecond pulsar
    J0218+4232.}
    \label{fig:systematic_study_1}
\end{figure*}

We briefly comment first on the results we obtain when incorporating the new averaging for the inner emission. 
Results are in general slightly  improved in comparison to \cite{systematic_2019} and to \cite{0218_analysis} for MSP J0218+4232:
We are able to successfully resemble the observational SEDs for the majority of the population studied. Details are shown in Table~\ref{tab:best_fit_parameters}.
A few exceptions, like J0357+3205 and J1826-1256, which in previous studies were not successfully fitted, are well fitted now, as shown in Figure \ref{fig:systematic_study_1}. 

The new averaging does not solve those few cases, like J2055+2539 and J0614+1807, which in \cite{systematic_2019} were not fitted well with one set of parameters.
As in the previous reference, we have considered in those cases the possibility
of having two separate accelerating regions, with equal values of $(E_\parallel,b)$ and different values of the geometrically related properties $(x_0,N_0)$ could deal well with the SED. 
As before, we find that this is the case and good models are obtained.
Then, whereas the averaging of the inner emission represents an overall 
improvement both in the physical description and in the ability of the model to fit pulsar spectra, it does not 
provide a disruptive change to our earlier results.

\begin{table*}
	\centering
	\caption{Results of the spectral fitting of the high-energy pulsar population with the minimal model ($x_{lp.in}/R_{lc}$ fixed, and averaging the spectra in the inner region).
	Best-fit values of the three model parameters ($\log E_{\parallel}$, $\log (x_0/R_{lc}), b$) are shown, together with the resulting value of $\overline{\chi^2}$ for each fit and the local strength of the magnetic field along the acceleration region $B$ at the start of the accelerating region.
	The two horizontal black lines divide the table in three sections.
	Pulsars in the upper section are those with $\overline{\chi^2}<2$.
	Pulsar in the middle section have $2 < \overline{\chi^2} < 3$.
	Finally, pulsars in the lower section have $\overline{\chi^2}>3$, and their fits are definitely not good.
	For the latter,
	errors are not given as the model needs two sets of parameters $(x_0/R_{lc},N_0)$
	to realize a good description.
	Pulsars with a star have fitting parameters violating the condition $E_\parallel < B$. Two of the cases were providing bad fits, 
	and are treated next. In the two other cases, an introduction of the condition constraints lead to the second-best model quoted.
	}
	\label{tab:best_fit_parameters}
	\begin{tabular}{lcccccccc} 
		\hline \hline
		 Pulsar & $\log{E_{\parallel}}$ [V/m] & $\log{(x_0/R_{lc})}$ & $b$ & $\log{N_0}$ & $\overline{\chi^2}$ & Local $B$ [G] & $B_{lc}^{b=3}$ [G] & $x_{lp.in}/R_{lc}$ \\
		 \hline
		J0007+7303  & $8.42^{+0.01}_{-0.01}$ & $-2.86^{+0.01}_{-0.01}$ & $2.86^{+0.02}_{-0.02}$ & $33.75$ & $1.12$ & $1.29 \times 10^5$ & $6.28 \times 10^3$ & $6.63 \times 10^{-6}$ \\ 
		J0205+6449  & $8.88^{+0.01}_{-0.01}$ & $-3.03^{+0.01}_{-0.01}$ & $3.02^{+0.02}_{-0.02}$ & $33.20$ &  $1.60$ & $1.66 \times 10^6$ & $2.32 \times 10^5$ & $3.19 \times 10^{-5}$ \\ 
		J0218+4232* & $10.80^{+0.01}_{-0.01}$ & $-4.06^{+0.01}_{-0.01}$ & $3.78^{+0.09}_{-0.08}$ & $32.93$ &  $0.76$ & $1.34 \times 10^6$ & $6.32 \times 10^5$ & $9.03 \times 10^{-6}$ \\
		 & $10.75^{+0.01}_{-0.02}$ & $-4.00^{+0.01}_{-0.02}$ & $3.59^{+0.11}_{-0.51}$ & $32.80$ & $0.86$ & $1.84 \times 10^{6}$ & " & " \\
		J0357+3205 & $6.66^{+0.04}_{-0.03}$ & $-0.59^{+0.06}_{-0.05}$ & $1.76^{+0.07}_{-0.08}$ & $31.71$ &  $0.80$ & $2.29 \times 10^7$ & $5.13 \times 10^2$ & $4.72 \times 10^{-5}$ \\
		J0633+1746 & $7.78^{+0.01}_{-0.01}$ & $-2.09^{+0.01}_{-0.01}$ & $2.56^{+0.02}_{-0.01}$ & $32.29$ &  $1.06$ & $3.00 \times 10^5$ & $2.26 \times 10^3$ & $8.84 \times 10^{-6}$ \\
		J0659+1414 & $8.84^{+0.01}_{-0.01}$ & $-3.90^{+0.01}_{-0.01}$ & $2.86^{+0.02}_{-0.02}$ & $33.57$ &  $1.16$ & $3.09 \times 10^4$ & $1.50 \times 10^3$ & $5.45 \times 10^{-6}$ \\
		J0751+1807* & $12.47^{+0.03}_{-0.03}$ & $-5.73^{+0.02}_{-0.02}$ &  $4.36^{+0.05}_{-0.05}$ & $32.61$ &  $1.63$ & $3.31 \times 10^4$ & $7.28 \times 10^4$ & $6.02 \times 10^{-6}$ \\   
		 & $10.46^{+0.06}_{-0.04}$ & $-3.74^{+0.04}_{-0.03}$ & $2.76^{+0.33}_{-0.51}$ & $31.12$ & $8.11$ & $9.68 \times 10^{5}$ & " & " \\
		J1357\textminus6429 & $8.29^{+0.03}_{-0.04}$ & $-2.92^{+0.03}_{-0.03}$ & $3.13^{+0.04}_{-0.05}$ & $34.35$ &  $1.08$ & $1.19 \times 10^5$ & $3.14 \times 10^4$ & $1.26 \times 10^{-5}$ \\
		J1420\textminus6048 & $8.61^{+0.03}_{-0.03}$ & $-2.62^{+0.03}_{-0.03}$ & $2.70^{+0.08}_{-0.08}$ & $33.80$ & $0.84$ & $5.27 \times 10^6$ & $1.40 \times 10^5$ & $3.07 \times 10^{-5}$ \\
		J1513\textminus5908 & $7.97^{+0.01}_{-0.01}$ & $-3.14^{+0.01}_{-0.01}$ & $3.66^{+0.02}_{-0.02}$ & $37.87$ &  $1.54$ & $1.33 \times 10^4$ & $8.19 \times 10^4$ & $1.39 \times 10^{-5}$ \\
		J1718\textminus3825 & $8.21^{+0.02}_{-0.02}$ & $-2.54^{+0.02}_{-0.02}$ & $3.19^{+0.03}_{-0.04}$ & $34.53$ &  $0.66$ & $1.30 \times 10^5$ & $4.44 \times 10^4$ & $2.81 \times 10^{-5}$ \\
		J1747\textminus2958 & $8.64^{+0.01}_{-0.01}$ & $-3.04^{+0.01}_{-0.01}$ & $3.01^{+0.02}_{-0.02}$ & $34.83$ &  $1.17$ & $3.62 \times 10^5$ & $4.75 \times 10^4$ & $2.12 \times 10^{-5}$ \\
		J1801\textminus2451 & $8.55^{+0.04}_{-0.06}$ & $-3.06^{+0.03}_{-0.05}$ & $3.07^{+0.05}_{-0.06}$ & $35.18$ &  $1.56$ & $2.07 \times 10^5$ & $3.81 \times 10^4$ & $1.68 \times 10^{-5}$ \\
		J1809\textminus2332 & $8.33^{+0.01}_{-0.01}$ & $-2.70^{+0.02}_{-0.01}$ & $2.96^{+0.05}_{-0.06}$ & $33.61$ &  $0.32$ & $1.31 \times 10^5$ & $1.32 \times 10^4$ & $1.43 \times 10^{-5}$ \\
		J1826\textminus1256 & $8.27^{+0.01}_{-0.01}$ & $-2.61^{+0.01}_{-0.01}$ & $3.13^{+0.01}_{-0.01}$ & $34.09$ &  $1.30$ & $1.95 \times 10^5$ & $5.08 \times 10^4$ & $1.90 \times 10^{-5}$ \\
		 J1833\textminus1034 & $9.58^{+0.02}_{-0.02}$ & $-3.84^{+0.02}_{-0.02}$ & $3.35^{+0.03}_{-0.03}$ & $35.27$ & $0.91$ & $3.96 \times 10^5$ & $2.78 \times 10^5$ & $3.39 \times 10^{-5}$ \\
		J1838\textminus0537 & $8.41^{+0.03}_{-0.03}$ & $-2.76^{+0.03}_{-0.03}$ & $3.16^{+0.04}_{-0.04}$ & $33.80$ & $1.58$ & $1.55 \times 10^5$ & $4.96 \times 10^4$ & $1.44 \times 10^{-5}$ \\
		J1846\textminus0258 & $6.09^{+0.04}_{-0.05}$ & $-1.12^{+0.05}_{-0.08}$ & $2.78^{+0.05}_{-0.04}$ & $34.66$ &  $1.22$ & $8.70 \times 10^5$ & $2.58 \times 10^4$ & $6.42 \times 10^{-6}$ \\
		J2021+3651& $8.46^{+0.01}_{-0.01}$ & $-2.78^{+0.01}_{-0.01}$ & $3.37^{+0.01}_{-0.01}$ & $34.50$ &  $1.10$ & $5.39 \times 10^4$ & $5.26 \times 10^4$ & $2.02 \times 10^{-5}$ \\
		 J2021+4026 & $8.10^{+0.01}_{-0.01}$ & $-2.70^{+0.01}_{-0.01}$ & $3.14^{+0.06}_{-0.04}$ & $34.93$ &  $0.95$ & $1.23 \times 10^4$ & $3.78 \times 10^3$ & $7.90 \times 10^{-6}$ \\
		 J2022+3842 & $8.58^{+0.03}_{-0.04}$ & $-2.84^{+0.03}_{-0.03}$ & $3.23^{+0.06}_{-0.08}$ & $34.99$ &  $0.44$ & $9.04 \times 10^5$ & $3.33 \times 10^5$ & $4.32 \times 10^{-5}$ \\
		 J2030+4415 & $8.83^{+0.03}_{-0.03}$ & $-3.49^{+0.03}_{-0.02}$ & $2.90^{+0.03}_{-0.03}$ & $33.75$ &  $0.52$ & $2.97 \times 10^4$ & $1.93 \times 10^3$ & $9.23 \times 10^{-6}$ \\
		 \hline
		 J0835\textminus4510 & $8.37^{+0.01}_{-0.01}$ & $-2.58^{+0.01}_{-0.01}$ & $3.34^{+0.03}_{-0.02}$ & $33.73$ &  $2.13$ & $1.12 \times 10^5$ & $8.73 \times 10^4$ & $2.35 \times 10^{-5}$ \\
		 J1813\textminus1246 & $8.58^{+0.01}_{-0.01}$ & $-2.66^{+0.01}_{-0.01}$ & $2.95^{+0.03}_{-0.03}$ & $34.00$ &  $2.13$ & $1.58 \times 10^6$ & $1.54 \times 10^5$ & $4.36 \times 10^{-5}$ \\
		 J1836+5925 & $7.72^{+0.01}_{-0.01}$ & $-1.93^{+0.01}_{-0.01}$ & $2.55^{+0.04}_{-0.03}$ & $31.79$ &  $2.47$ & $2.23 \times 10^5$ & $1.83 \times 10^3$ & $1.21 \times 10^{-5}$ \\
		J2055+2539 & $7.13^{+0.03}_{-0.03}$ & $-1.37^{+0.03}_{-0.02}$ & $1.51^{+0.03}_{-0.03}$ & $31.03$ &  $2.43$ & $1.07 \times 10^8$ & $6.53 \times 10^2$ & $6.56 \times 10^{-6}$ \\
		J2229+6114 & $8.63^{+0.01}_{-0.01}$ & $-2.50^{+0.01}_{-0.01}$ &  $2.65^{+0.03}_{-0.03}$ & $33.27$ &  $2.27$ & $1.18 \times 10^7$ & $2.72 \times 10^5$ & $4.06 \times 10^{-5}$ \\
		\hline
		J0437\textminus4715* & $10.80$ & $-4.39$ & $2.87$ & $31.50$ &  $5.64$ & $6.28 \times 10^5$ & $5.60 \times 10^4$ & $3.64 \times 10^{-5}$ \\
		J0614\textminus3329 & $9.47$ & $-2.50$ & $2.95$ & $30.01$ &  $15.27$ & $1.24 \times 10^6$ & $1.41 \times 10^5$ & $6.65 \times 10^{-6}$ \\
		J1057\textminus5226 &  $7.28$ & $-1.21$ & $1.58$ & $30.86$ &  $7.23$ & $1.31 \times 10^8$ & $2.60 \times 10^3$ & $1.06 \times 10^{-5}$
		\\
		J1124\textminus5916 & $9.08$ & $-3.62$ & $3.23$ & $35.43$ &  $3.89$ & $1.62 \times 10^5$ & $7.55 \times 10^4$ & $1.55 \times 10^{-5}$ \\
		J1231\textminus1411 & $8.87$ & $-1.79$ & $3.03$ & $29.13$ &  $53.40$ & $7.87 \times 10^5$ & $1.04 \times 10^5$ & $5.70 \times 10^{-6}$ \\
		J1709\textminus4429& $8.68$ & $-2.94$ & $3.30$ & $35.18$ &  $4.27$ & $8.53 \times 10^4$ & $5.34 \times 10^4$ & $2.04 \times 10^{-5}$ \\
		J1741\textminus2054 & $7.04$ & $-1.61$ & $1.63$ & $31.14$ &  $6.24$ & $7.24 \times 10^7$ & $6.98 \times 10^2$ & $5.07 \times 10^{-6}$ \\
		J1952+3252*  & $10.75$ & $-4.80$ & $3.39$ & $34.16$ &  $10.09$ & $2.04 \times 10^5$ & $1.45 \times 10^5$ & $5.32 \times 10^{-6}$ \\
		J2124\textminus3358 & $8.61$ & $-1.42$ & $2.79$ & $29.16$ & $34.23$ & $6.70 \times 10^5$ & $4.96 \times 10^4$ & $4.25 \times 10^{-5}$ \\
		\hline
		\hline
	\end{tabular}
\end{table*}

In some cases the $E_\parallel$ listed in Table \ref{tab:best_fit_parameters} may be larger than the local $B$, which is not physical.  However,
in the spirit of the discussion of possible violations of the condition (see  
\cite{Uzdensky2003}), we comment these cases explicitly.
The violation of this condition occurs in four cases: J0218+4232, J0751+1807, J04337-4715 and J1952+3252. However, the two latter cases were not well fitted with the minimal model and results for them are thus unaffected, since when we consider next a free $x_{lp.in}$ this constraint is fulfilled.
For the first two cases, i.e., J0218+4232, J0751+1807, we have performed the fitting again adding this limitation directly in the algorithm. For J0218+4232 we still find a good fit which fulfills the $E_\parallel < B$ criterion (new best-fit parameters are commented on in Table \ref{tab:best_fit_parameters}). Changes are minimal, from $(E_\parallel, x_0, b: \chi^2)$=(10.79, -4.06, 3.77: 0.76) to (10.77, -4.00, 3.59: 0.85). We have found no sufficiently good fit that fulfills this condition for J0751+1807.

\subsection{Fitting the spatial scale $x_{lp.in}$}\label{fit_x_lpin}

\begin{table*}
	\centering
	\caption{F-test for the pulsars whose fit by the minimal model gives $\overline{\chi^2}>2$.
	This model (in which the value of $x_{lp.in}/R_{lc}$ is given by Eq. \eqref{eq:x_birth}) is compared with the model with $x_{lp.in}/R_{lc}$ free. 
	D.O.F is the number of degrees of freedom of the model, defined as the difference between $N_{bin}$ and the number of model parameters. 
	Critical values are defined by the pair of numbers of degrees of freedom of the two models that are compared.
	To define such values a 1 \% confidence level is chosen.
	Pulsars on the upper section are those shown on Figure \ref{fig:xbirth_scan_1}, while those on the  lower section are those shown on Figure \ref{fig:xbirth_scan_2}. }
	\label{tab:f_tests}
	\begin{tabular}{ccccccccccc} 
		\hline \hline
		 Pulsar & Model & $\log{E_{\parallel}}$ [V/m] & $\log{(x_0/R_{lc})}$ & $b$ & $\log{N_0}$ & $\log{x_{lp.in}/R_{lc}}$ & $\overline{\chi^2}$ & D.O.F. & F-value & Critical value \\ 
		\hline
		 \multirow{2}{*}{J0835\textminus4510} & Minimal & $8.37^{+0.01}_{-0.01}$ & $-2.58^{+0.01}_{-0.01}$ & $3.34^{+0.03}_{-0.02}$ & $33.73$ & $-4.63$ & $2.13$ & $37$ & \multirow{2}{*}{$37.71$} & \multirow{2}{*}{$2.20$} \\%
		 & x$_{lp.in}$ free & $8.27^{+0.01}_{-0.01}$ & $-2.41^{+0.01}_{-0.01}$ & $3.06^{+0.03}_{-0.03}$ & $31.65$ & $-6.78^{+0.97}_{-...}$ &  $1.07$ & $36$ &  & \\
		 \multirow{2}{*}{J1813\textminus1246} & Minimal & $8.58^{+0.01}_{-0.01}$ & $-2.66^{+0.01}_{-0.01}$ & $2.95^{+0.03}_{-0.03}$ & $34.00$ & $-4.36$ & $2.13$ & $31$ & \multirow{2}{*}{$23.12$} & \multirow{2}{*}{$2.38$} \\%
		 & x$_{lp.in}$ free & $10.71^{+0.02}_{-0.02}$ & $-4.74^{+0.02}_{-0.02}$ & $2.64^{+0.01}_{-0.01}$ & $31.74$ & $-6.02^{+0.06}_{-0.02}$ &  $1.25$ & $30$ &  & \\
		 \multirow{2}{*}{J1836+5925} & Minimal & $7.72^{+0.01}_{-0.01}$ & $-1.93^{+0.01}_{-0.01}$ & $2.55^{+0.04}_{-0.03}$ & $31.79$ & $-4.92$ & $2.47$ & $16$ & \multirow{2}{*}{$26.44$} & \multirow{2}{*}{$3.49$} \\%
		 & x$_{lp.in}$ free &  $7.73^{+0.01}_{-0.01}$ & $-1.96^{+0.01}_{-0.01}$ & $2.78^{+0.02}_{-0.02}$ & $33.04$ & $-3.73^{+0.08}_{-0.11}$ &  $0.95$ & $15$ &  & \\
		 \multirow{2}{*}{J2055+2539} & Minimal & $7.13^{+0.03}_{-0.03}$ & $-1.37^{+0.03}_{-0.02}$ & $1.51^{+0.03}_{-0.03}$ & $31.03$ & $-5.18$ & $2.43$ & $27$ & \multirow{2}{*}{$30.52$} & \multirow{2}{*}{$2.54$} \\%
		 & x$_{lp.in}$ free &  $6.97^{+0.04}_{-0.05}$ & $-0.85^{+0.03}_{-0.04}$ & $1.47^{+0.03}_{-0.02}$ & $31.69$ & $-4.12^{+0.08}_{-0.05}$ &  $1.16$ & $26$ &  & \\
		 \multirow{2}{*}{J2229+6114} & Minimal & $8.63^{+0.01}_{-0.01}$ & $-2.50^{+0.01}_{-0.01}$ &  $2.65^{+0.03}_{-0.03}$ & $33.27$ & $-4.39$ & $2.28$ & $25$ & \multirow{2}{*}{$1.06$} & \multirow{2}{*}{$2.64$} \\%
		 & x$_{lp.in}$ free &  $8.91^{+0.01}_{-0.01}$ & $-3.00^{+0.01}_{-0.01}$ & $3.19^{+0.02}_{-0.02}$ & $33.96$ &  $-4.92^{+0.05}_{-0.06}$ &  $2.27$ & $24$ &  & \\
		\hline
		 \multirow{2}{*}{J0437\textminus4715} & Minimal & $10.80$ & $-4.39$ & $2.87$ & $31.50$ &  $-4.44$ & $5.64$ & $32$ & \multirow{2}{*}{$10.69$} & \multirow{2}{*}{$2.34$} \\%
		 & x$_{lp.in}$ free &  $9.19^{+0.01}_{-0.01}$ &  $-2.75^{+0.01}_{-0.01}$ &   $3.10^{+0.02}_{-0.02}$  & $32.92$ &  $-2.49^{+0.02}_{-0.02}$ &  $4.33$ & $31$ &  & \\
		 \multirow{2}{*}{J0614\textminus3329} & Minimal & $9.47$ & $-2.50$ & $2.95$ & $30.01$ & $-5.18$ & $15.27$ & $28$ & \multirow{2}{*}{$145.15$} & \multirow{2}{*}{$2.49$} \\%
		 & x$_{lp.in}$ free & $10.88^{+0.01}_{-0.01}$ &  $-4.05^{+0.01}_{-0.01}$  &  $5.09^{+0.02}_{-0.02}$  & $35.34$ & $-3.34^{+0.01}_{-0.01}$ &  $4.43$ & $27$ &  & \\ 
		\multirow{2}{*}{J1057\textminus5226} & Minimal &  $7.28$ & $-1.21$ & $1.58$ & $30.86$ &  $-4.97$ & $7.23$ & $13$ & \multirow{2}{*}{$1.77$} & \multirow{2}{*}{$4.10$} \\%
		 & x$_{lp.in}$ free &  $7.25^{+0.02}_{-0.02}$ &  $-1.11^{+0.03}_{-0.03}$  &  $1.57^{+0.02}_{-0.02}$  & $31.01$ & $-4.76^{+0.08}_{-0.05}$ &  $6.83$ & $12$ &  & \\
		 \multirow{2}{*}{J1124\textminus5916} & Minimal & $9.08$ & $-3.62$ & $3.23$ & $35.43$ & $-4.81$ & $3.89$ & $30$ & \multirow{2}{*}{$11.36$} & \multirow{2}{*}{$2.41$} \\%
		 & x$_{lp.in}$ free &  $8.03^{+0.01}_{-0.02}$ &  $-2.47^{+0.01}_{-0.01}$  &  $3.31^{+0.01}_{-0.01}$   & $36.40$ & $-2.90^{+0.04}_{-0.04}$ &  $2.89$ & $29$ &  & \\
		 \multirow{2}{*}{J1231\textminus1411} & Minimal & $8.87$ & $-1.79$ & $3.03$ & $29.13$ & $-5.24$ & $53.40$ & $28$ & \multirow{2}{*}{$0.81$} & \multirow{2}{*}{$2.49$} \\%
		 & x$_{lp.in}$ free & $8.94^{+0.02}_{-0.02}$ &  $-1.91^{+0.03}_{-0.02}$  &  $2.98^{+0.19}_{-0.15}$  & $31.04$ & $-3.26^{+0.22}_{-0.10}$ &  $53.77$ & $27$ &  & \\
		 \multirow{2}{*}{J1709\textminus4429} & Minimal & $8.68$ & $-2.94$ & $3.30$ & $35.18$ &  $-4.69$ & $4.27$ & $14$ & \multirow{2}{*}{$28.34$} & \multirow{2}{*}{$3.86$} \\%
		 & x$_{lp.in}$ free & $8.85^{+0.01}_{-0.01}$ & $-3.11^{+0.01}_{-0.01}$ &   $3.14^{+0.02}_{-0.02}$ & $34.14$ & $-5.81^{+0.09}_{-0.11}$ &  $1.45$ & $13$ &  & \\
		 \multirow{2}{*}{J1741\textminus2054} & Minimal & $7.04$ & $-1.61$ & $1.63$ & $31.14$ & $-5.30$ & $6.24$ & $38$ & \multirow{2}{*}{$46.74$} & \multirow{2}{*}{$2.17$} \\%
		 & x$_{lp.in}$ free &  $7.63^{+0.01}_{-0.01}$ &  $-2.40^{+0.01}_{-0.01}$  &  $2.70^{+0.01}_{-0.01}$  & $34.25$ & $-3.32^{+0.04}_{-0.04}$ &  $2.83$ & $37$ &  & \\
		 \multirow{2}{*}{J1952+3252} & Minimal & $10.75$ & $-4.80$ & $3.39$ & $34.16$ & $-5.28$ & $10.09$ & $25$ & \multirow{2}{*}{$8.95$} & \multirow{2}{*}{$2.64$} \\%
		 & x$_{lp.in}$ free &  $8.45^{+0.01}_{-0.01}$ &  $-2.42^{+0.01}_{-0.01}$  &  $3.49^{+0.01}_{-0.01}$  & $36.35$ & $-2.36^{+0.02}_{-0.02}$ &  $7.66$ & $24$ &  & \\
		 \multirow{2}{*}{J2124\textminus3358} & Minimal & $8.61$ & $-1.42$ & $2.79$ & $29.16$ & $-4.37$ & $34.23$ & $20$ & \multirow{2}{*}{$194.80$} & \multirow{2}{*}{$3.00$} \\
		 & x$_{lp.in}$ free & $8.98^{+0.02}_{-0.02}$  & $-2.16^{+0.01}_{-0.01}$  &  $4.27^{+0.02}_{-0.02}$ & $35.52$ & $-1.40^{+0.01}_{-0.01}$ &  $3.20$ & $19$ &  & \\
		 \hline
		\hline
	\end{tabular}
\end{table*}

\begin{figure*}
            \begin{subfigure}{.33\textwidth}
                \centering
                \includegraphics[width=1.0\linewidth]{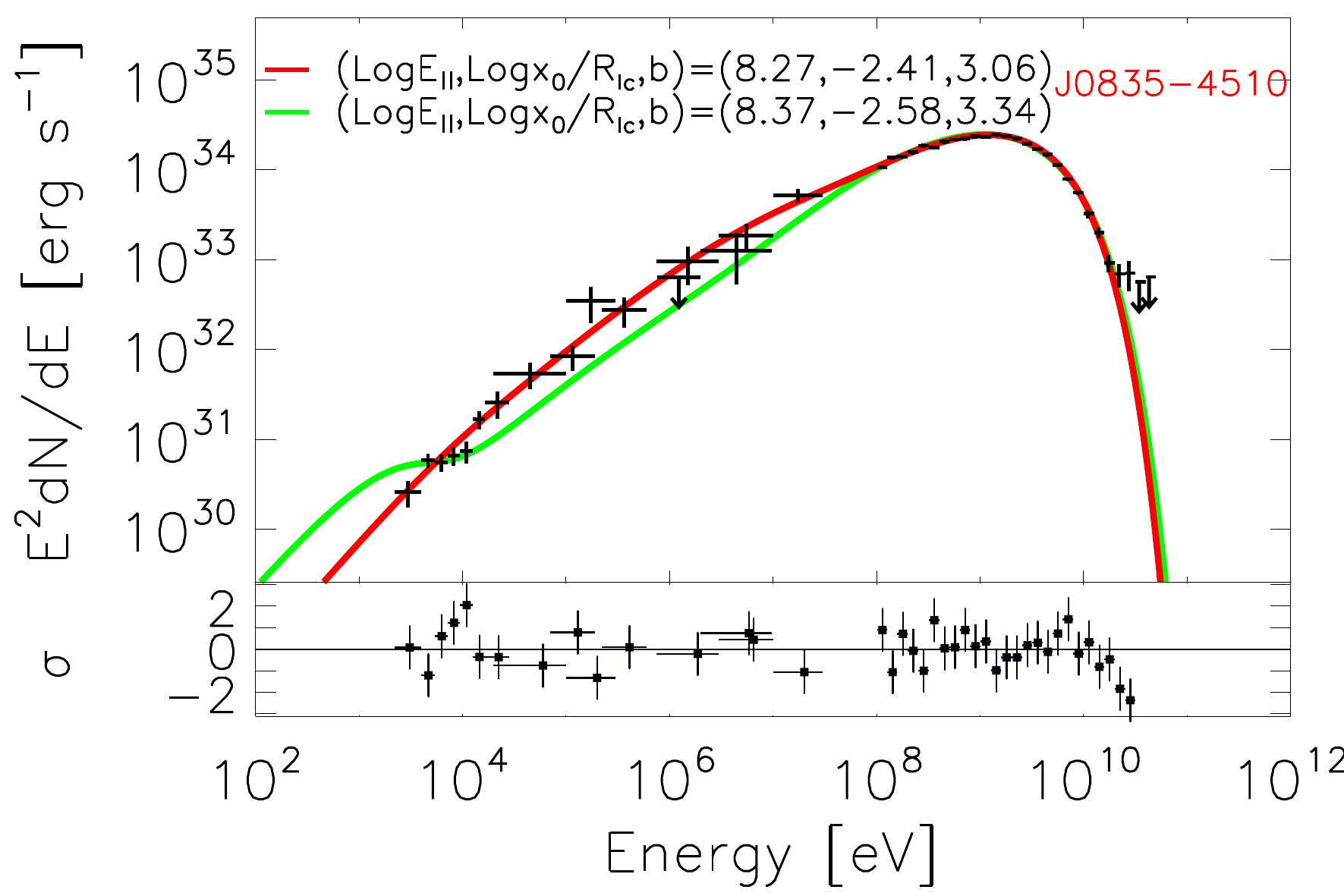}
                \label{fig:vela_xbirth_free}
            \end{subfigure}%
            \begin{subfigure}{.33\textwidth}
                \centering
                \includegraphics[width=1.0\linewidth]{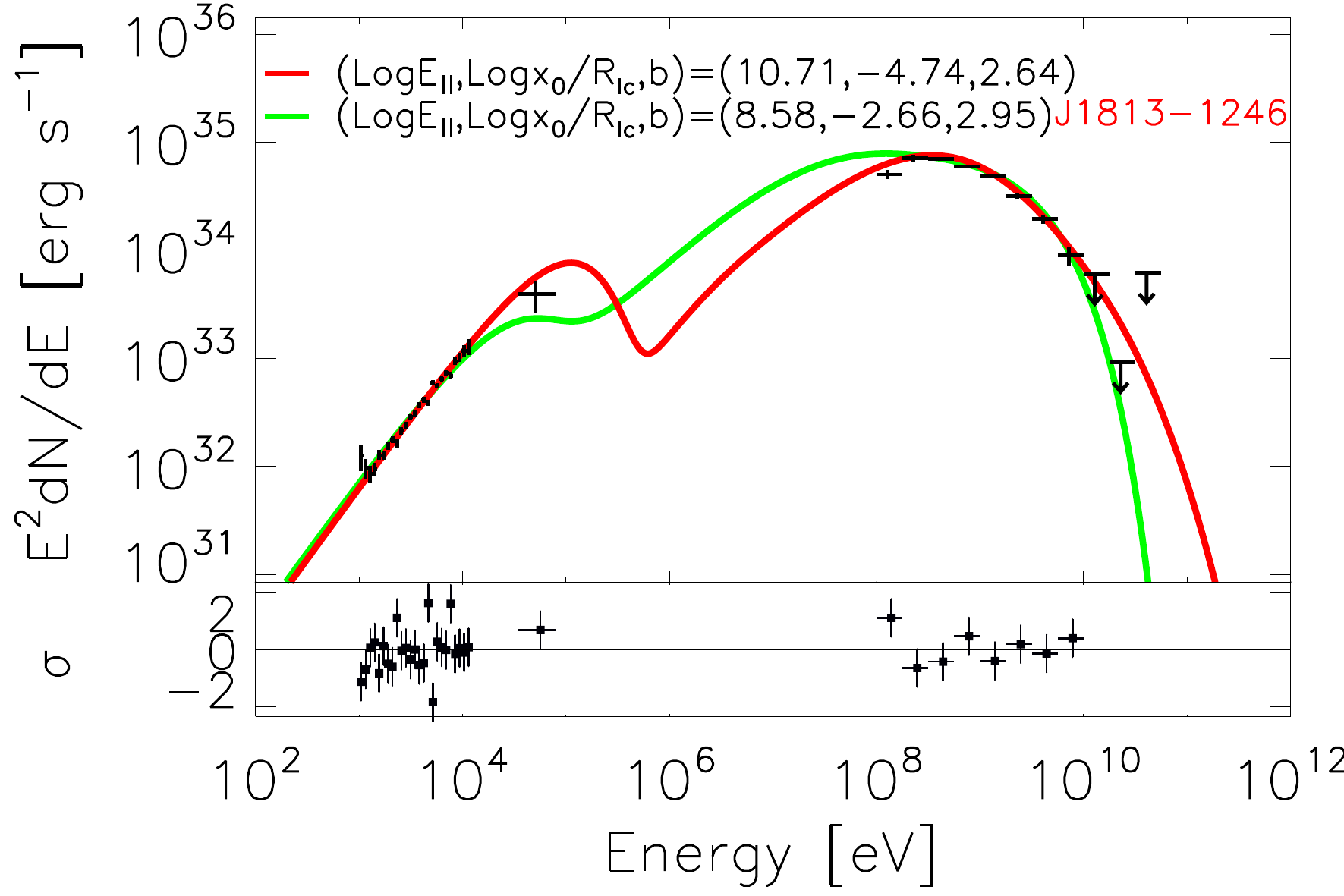}
                \label{fig:1813_xbirth_free}
            \end{subfigure}%
            \begin{subfigure}{.33\textwidth}
                \centering
                \includegraphics[width=1.0\linewidth]{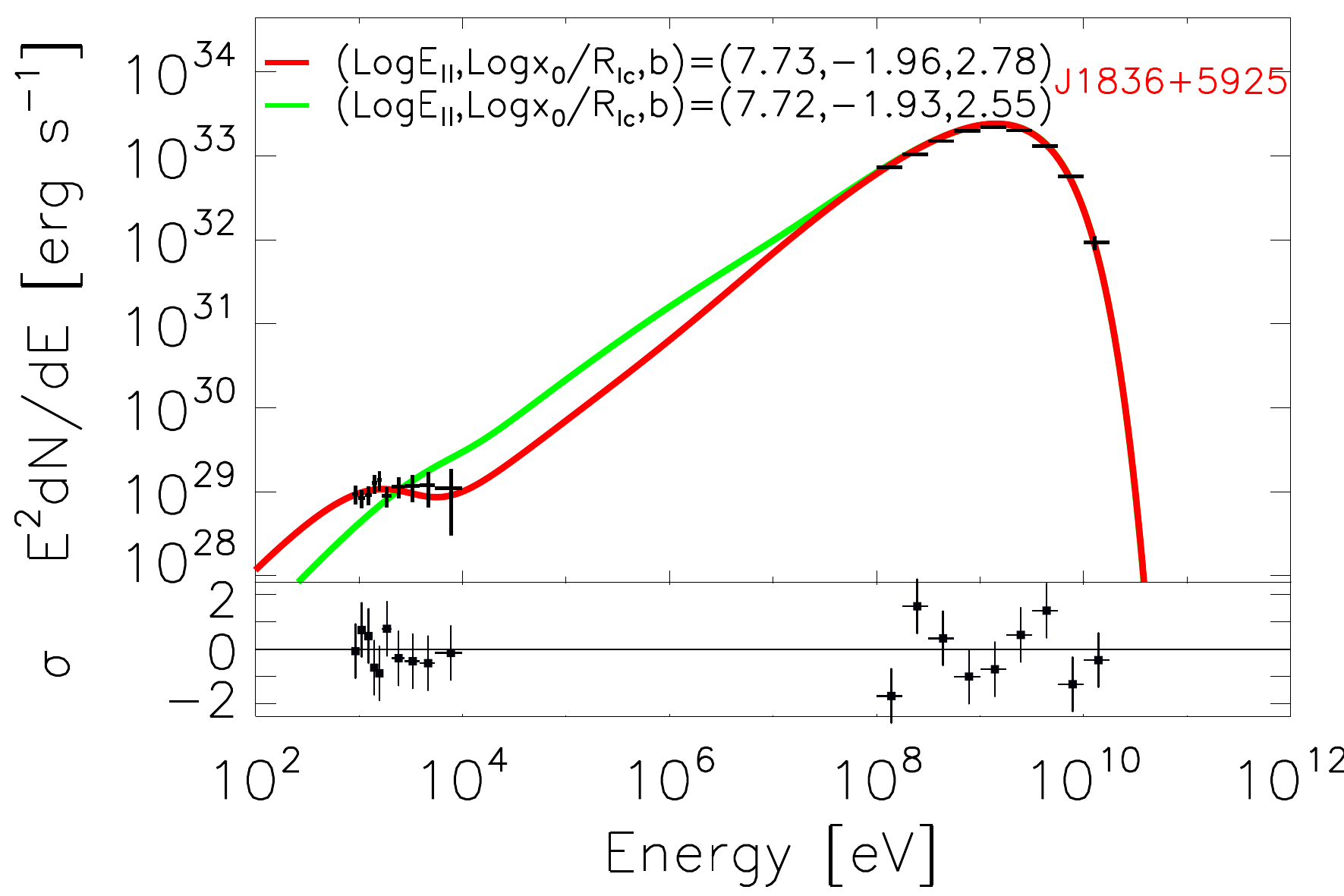}
                \label{fig:1836_xbirth_free}
            \end{subfigure}
            \centering
            \begin{subfigure}{.33\textwidth}
                \centering
                \includegraphics[width=1.0\linewidth]{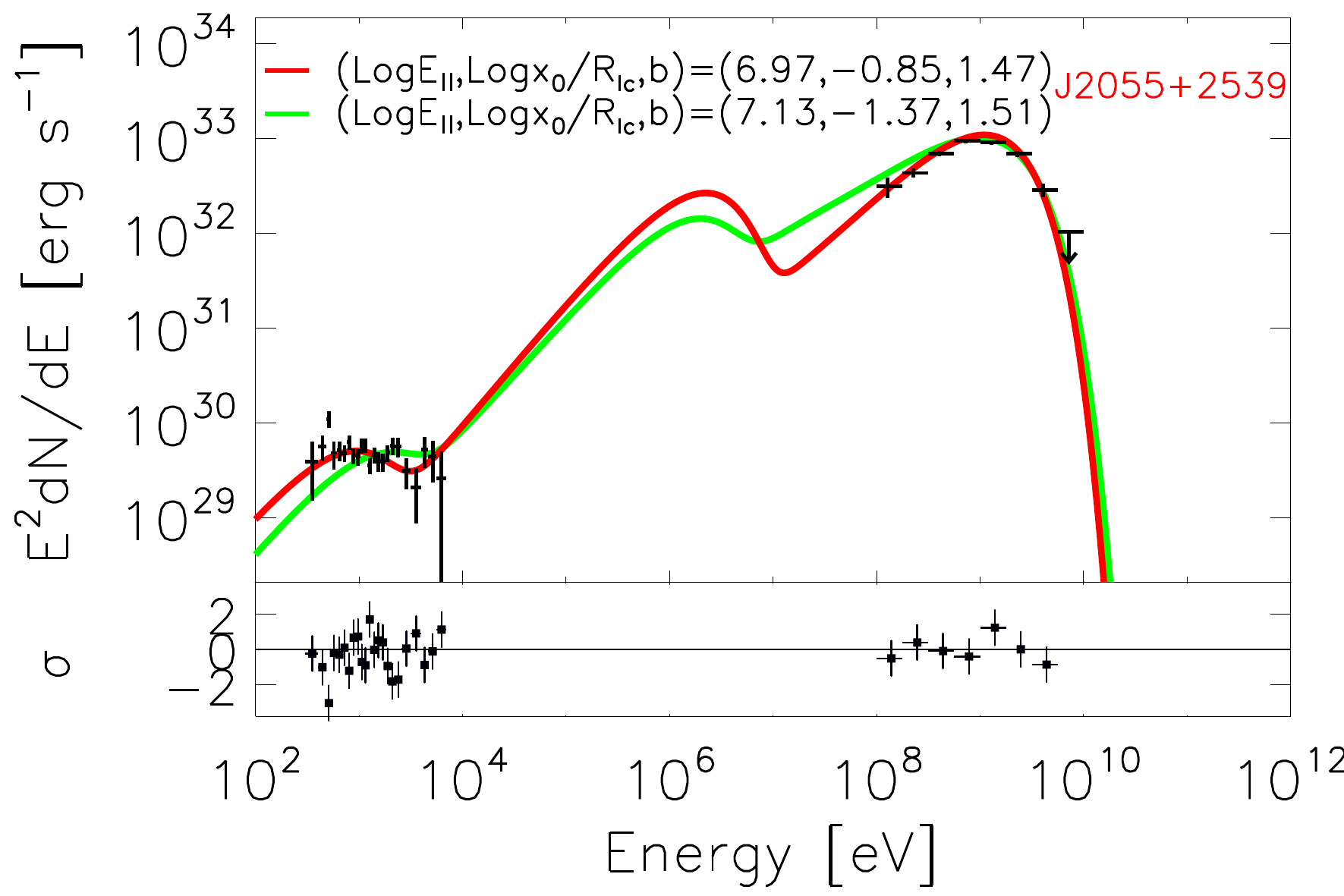}
                \label{fig:2055_xbirth_free}
            \end{subfigure}%
            \begin{subfigure}{.33\textwidth}
                \centering
                \includegraphics[width=1.0\linewidth]{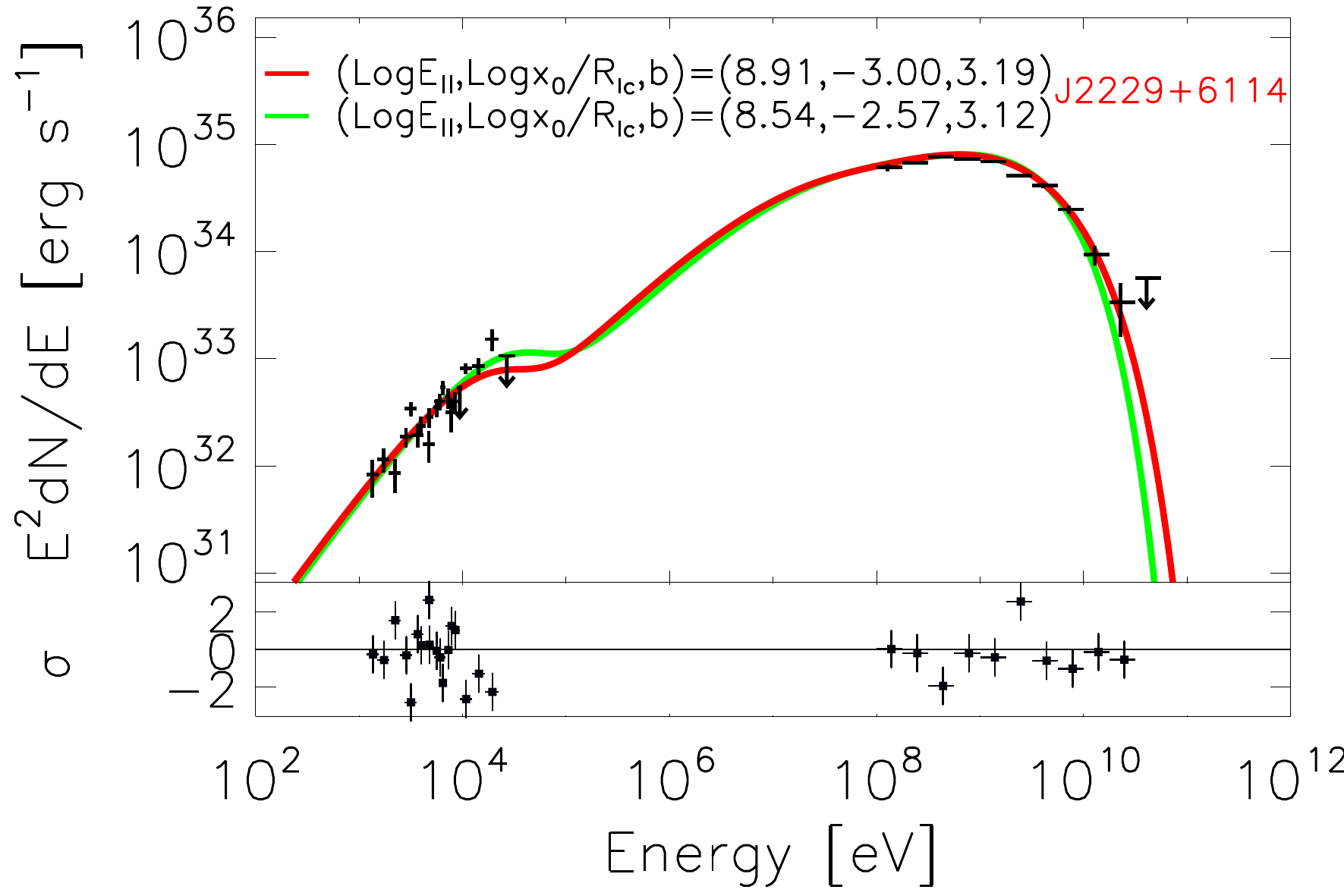}
                \label{fig:2229_xbirth_free}
            \end{subfigure}
            \caption{Comparison of the theoretical SEDs produced with the minimal model (green lines) and with the model having $x_{lp.in}/R_{lc}$ as a free parameter (red  lines). The residuals are those of the spectrum where $x_{lp.in}/R_{lc}$ is free. Pulsars shown here are those whose fit with the minimal model has $2 < \overline{\chi^2} < 3$.}
            \label{fig:xbirth_scan_1}
    \end{figure*}

In addition of the minimization for the three parameters described above, we implemented an overarching bisection method for $x_{lp.in}/R_{lc}$.
That is, for each evaluation within the bisection, an ($x_{lp.in}/R_{lc}$)-value is considered and we perform the \texttt{2+1+3} procedure explained in \S \ref{fitting_approach}, which in this case acts as the function to minimize.
The searching range for $x_{lp.in}/R_{lc}$ was  set at 
two orders of magnitude above and below,
\begin{equation}
    \frac{x_{lp.in}}{R_{lc}} \sim 10^{-1.0 \times \log_{10}{(R_{lc} [cm])} + 4.0}
    \label{eq:x_birth}
\end{equation}
since we find that this expression provides a good representation of the simplified treatment used for our earlier works as described in \S\ref{birth-treat}.
We shall refer to the case when we choose to fix $x_{lp.in}$ at the value given by Eq. \ref{eq:x_birth} as minimal model (i.e. three parameters with $x_{lp.in}$ fixed as explained above), and use that as a reference for a comparison with results arising with a ranging $x_{lp.in}$ (i.e., having four parameters in total).

For those pulsars in which the minimal model was providing a good fit (first panel of Table 1), considering $x_{lp.in}$ as a free parameter would in general lead to 
recover values close to the one assumed in the minimal scenario, confirming all such previous fits. We are thus particularly interested in those cases where the minimal model were not particularly good, those in the middle and lower sections of Table \ref{tab:best_fit_parameters}.
Table \ref{tab:f_tests} shows the results of the model with $x_{lp.in}/R_{lc}$ free and compares them with those for the minimal case for the latter.
The pulsars listed are those with $\overline{\chi^2}> 2$, and are divided between those with $2 < \overline{\chi^2} < 3$ and those with $\overline{\chi^2} > 3$.

Table \ref{tab:f_tests}  also shows the results of an F-test that compares the two models (the minimal model and the model with  $x_{lp.in}$ free) under the null hypothesis 
of no improvement when increasing the number of free parameters. 
The F-value is defined as
\begin{equation}
    F = \left[{\dfrac{RSS_1 - RSS_2}{n_1 - n_2}}\right] / \left[{\dfrac{RSS_2}{n_2}}\right],
    \label{eq:f_value}
\end{equation}
where $RSS$ is the residual sum of squares of the model and $n$ is the number of degrees of freedom of each model, defined as $N_{bin} - p$, with $p$ the number of free parameters. In our  case, $RSS$ is $\chi^2$. 
The F-value is then compared to a critical value of the F-distribution, which depends on the 
numbers of degrees of freedom of the models that are being compared and a desired confidence interval.
The critical value of a F-distribution is the value of the distribution for which, if the F-value of the comparison of models is larger than it, the null hypothesis is rejected (in our case: the model with free $x_{lp.in}$ is preferred).

\begin{figure*}
            \begin{subfigure}{.35\textwidth}
                \centering
                \includegraphics[width=1.0\linewidth]{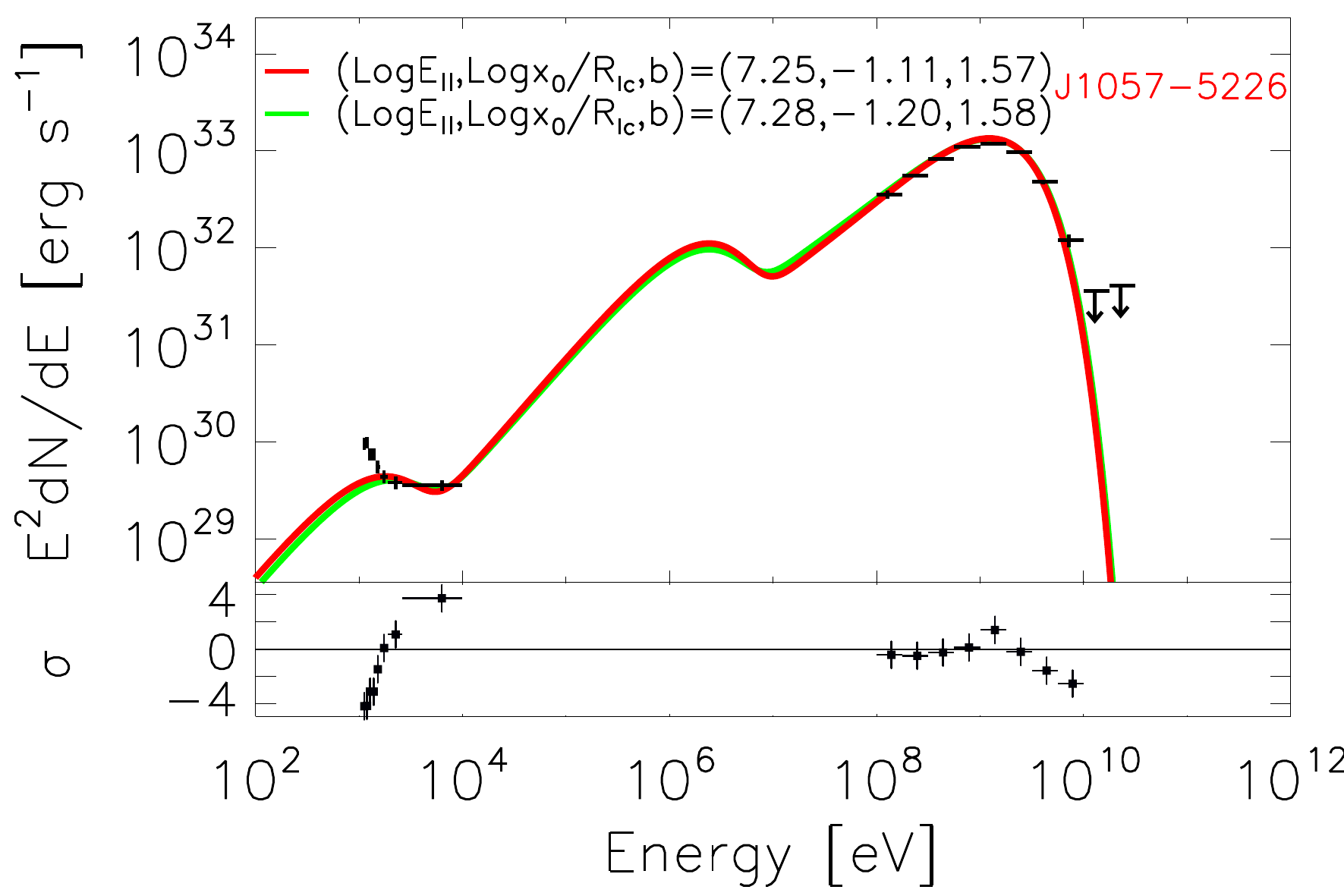}
                \label{fig:1057_xbirth_free}
            \end{subfigure}%
            \begin{subfigure}{.35\textwidth}
                \centering
                \includegraphics[width=1.0\linewidth]{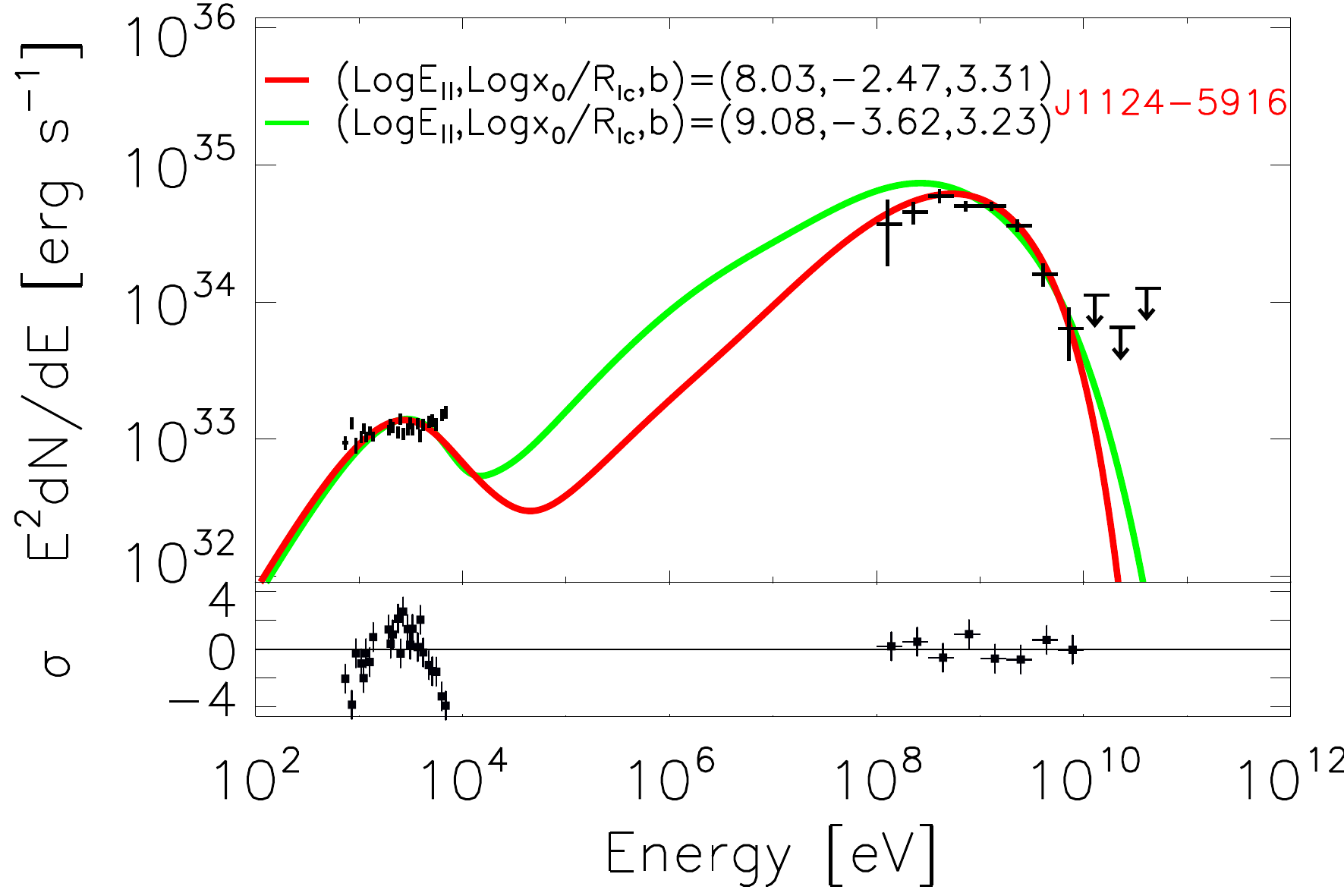}
                \label{fig:112b_xbirth_free}
            \end{subfigure}%
            \begin{subfigure}{.35\textwidth}
                \centering
                \includegraphics[width=1.0\linewidth]{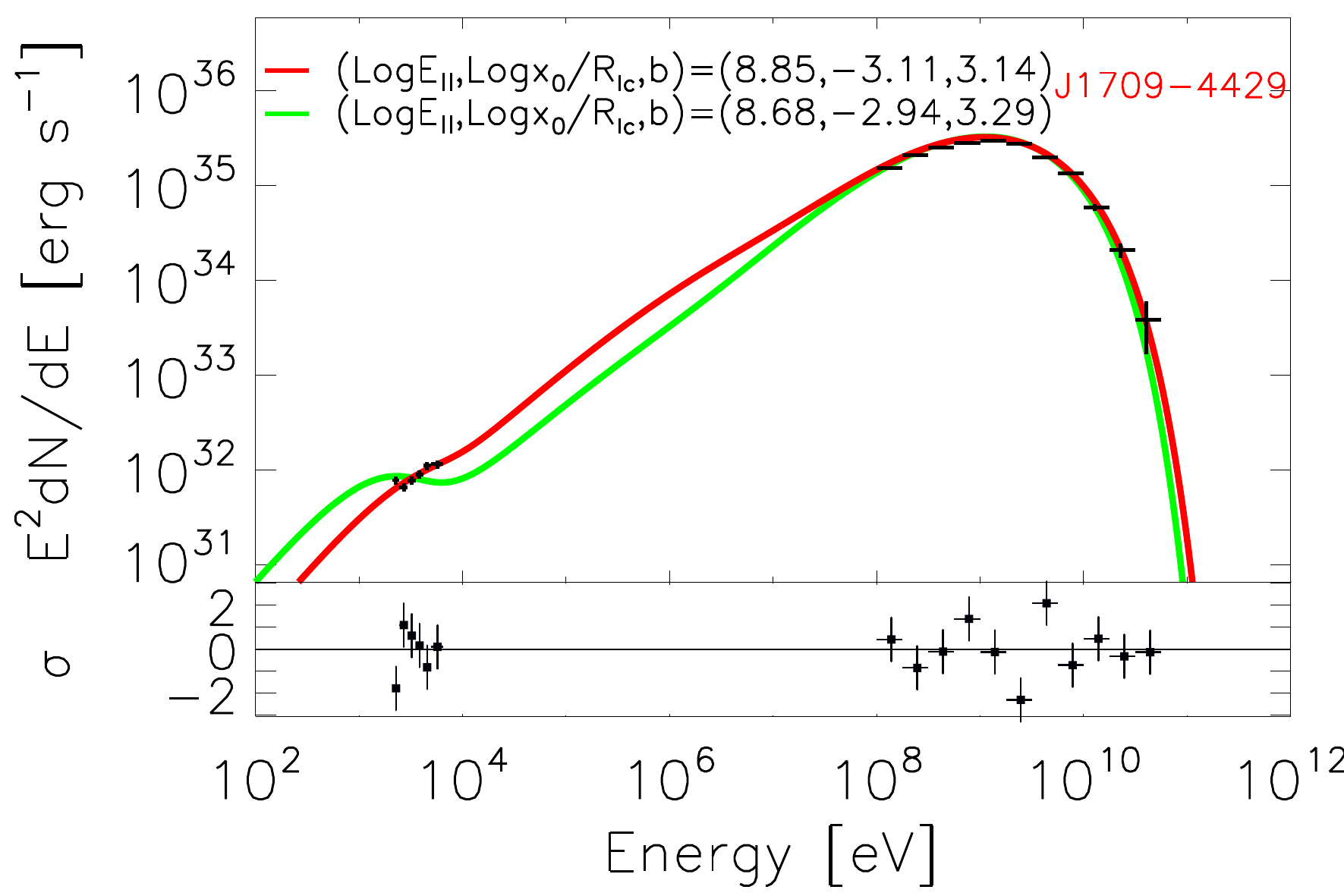}
                \label{fig:1709_xbirth_free}
            \end{subfigure}
            \begin{subfigure}{.35\textwidth}
                \centering
                \includegraphics[width=1.0\linewidth]{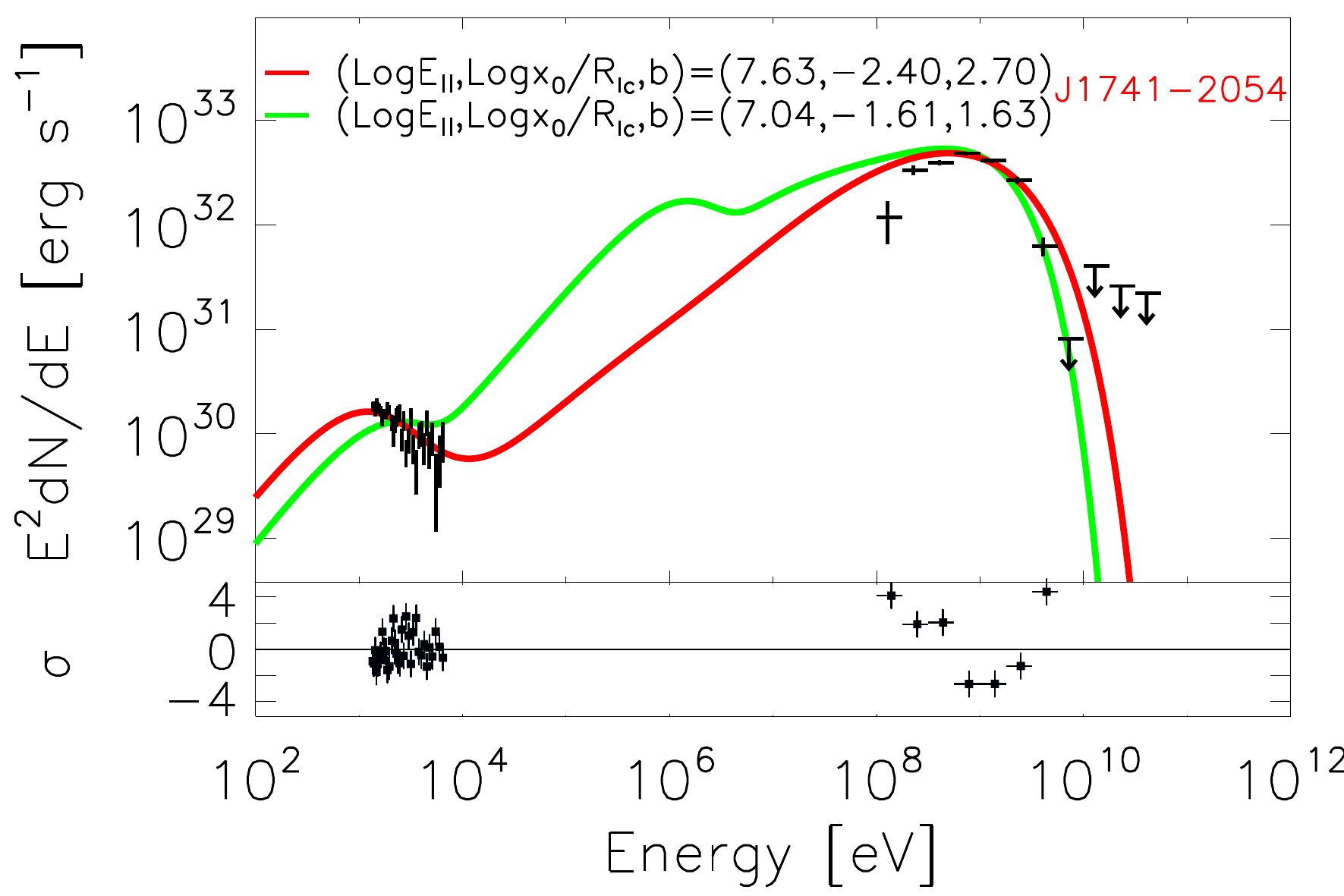}
                \label{fig:1741_xbirth_free}
            \end{subfigure}%
            \begin{subfigure}{.35\textwidth}
                \centering
                \includegraphics[width=1.0\linewidth]{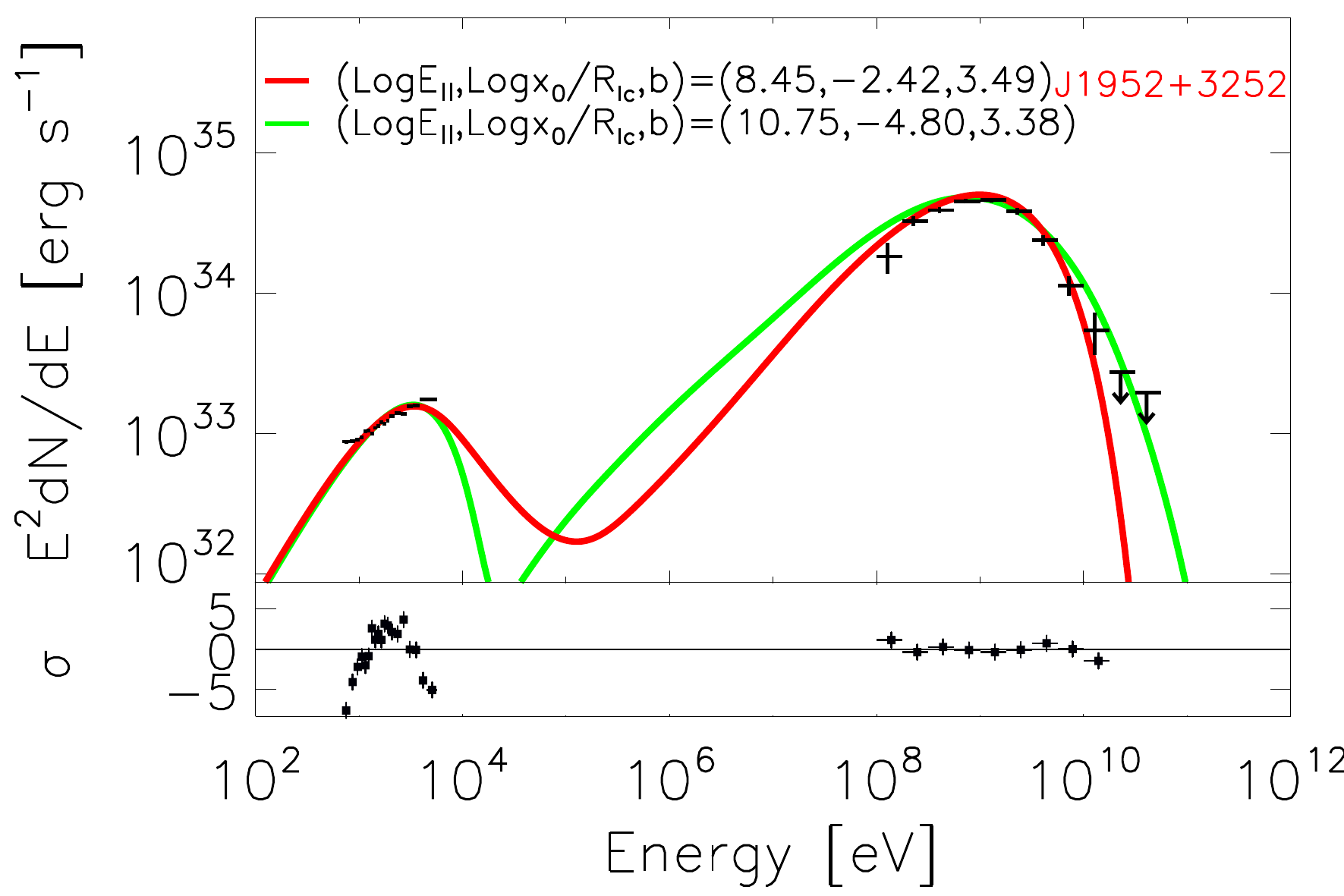}
                \label{fig:1952_xbirth_free}
            \end{subfigure}%
            \begin{subfigure}{.35\textwidth}
                \centering
                \includegraphics[width=1.0\linewidth]{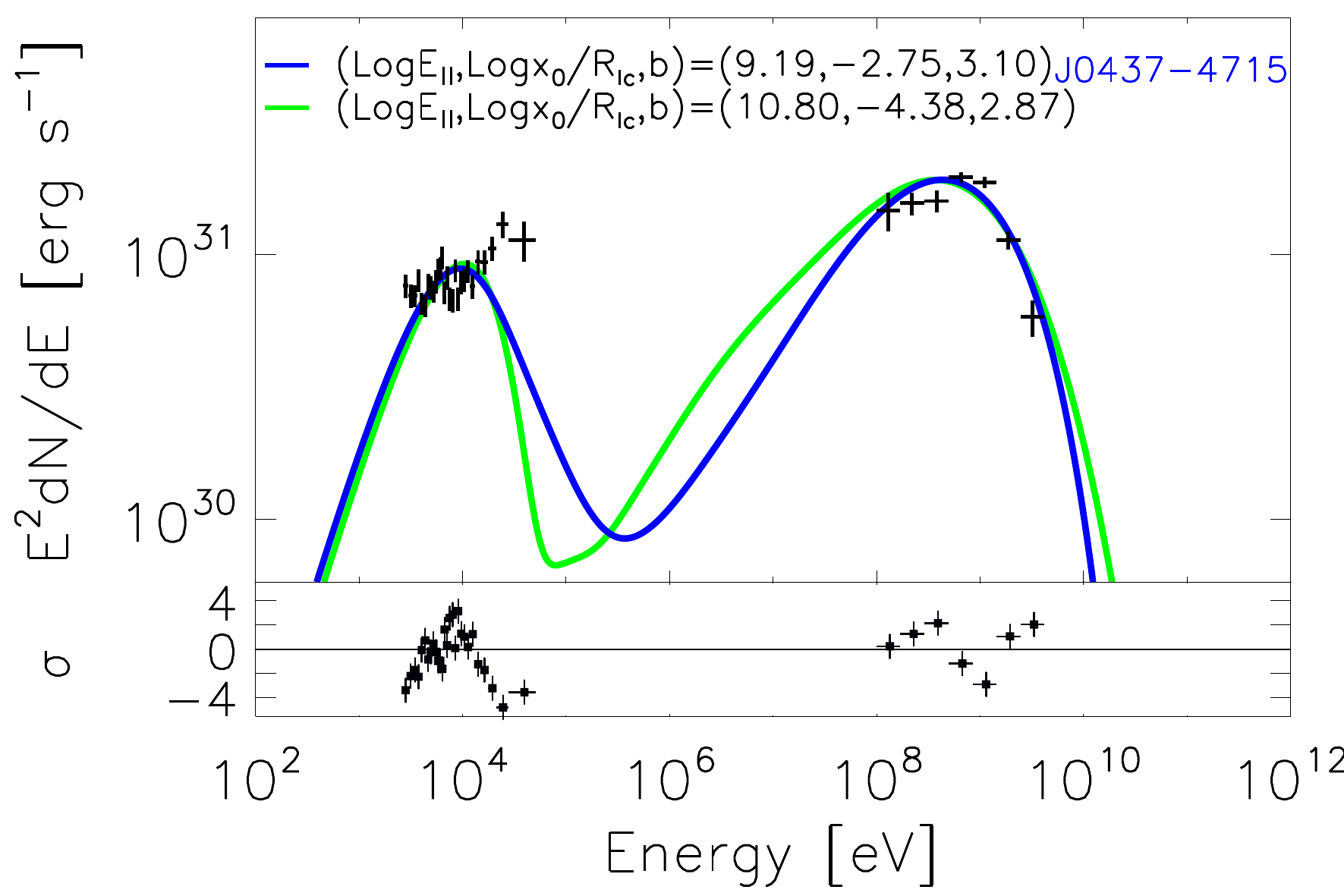}
                \label{fig:0437_xbirth_free}
            \end{subfigure}
            \begin{subfigure}{.35\textwidth}
                \centering
                \includegraphics[width=1.0\linewidth]{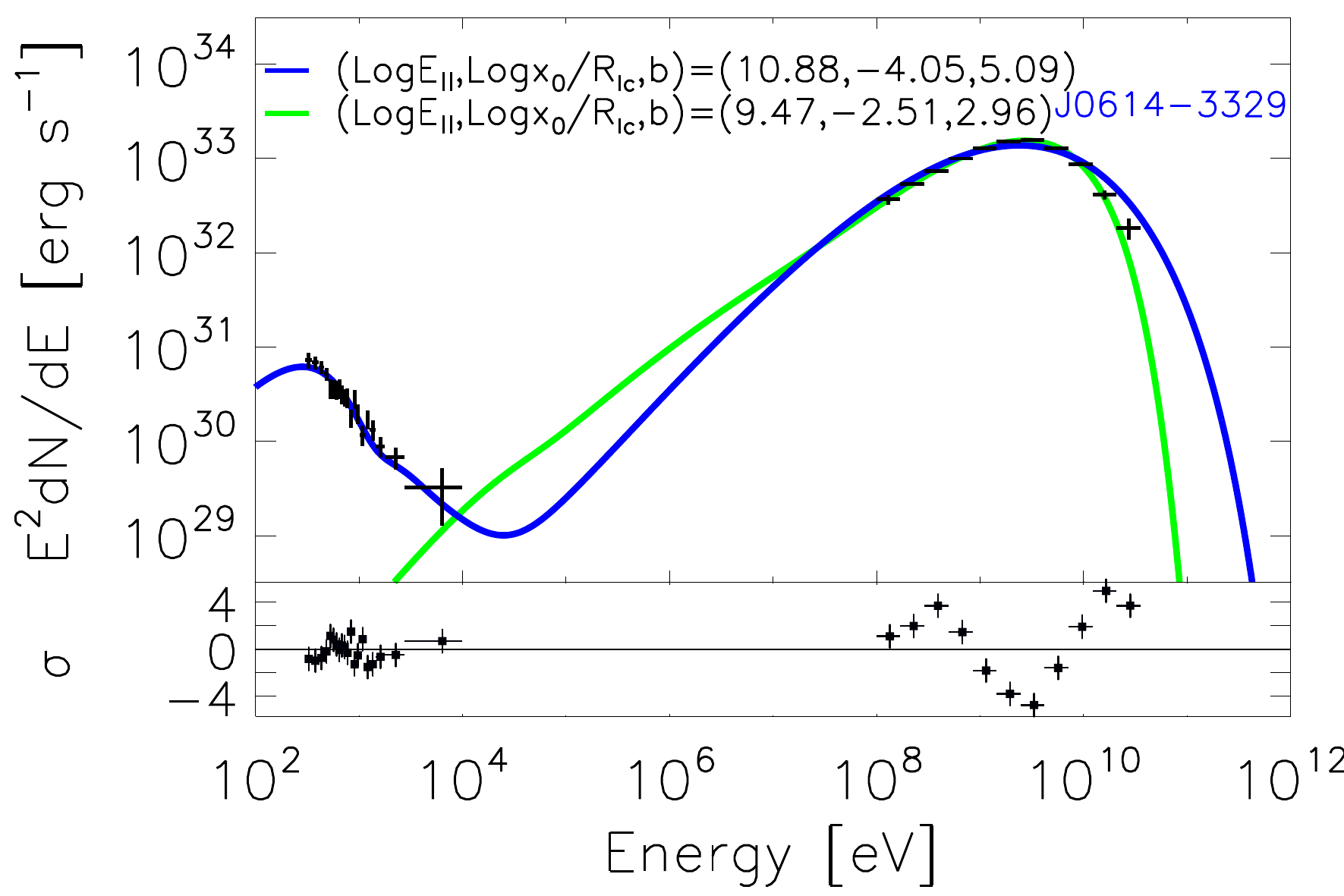}
                \label{fig:0614_xbirth_free}
            \end{subfigure}%
            \begin{subfigure}{.35\textwidth}
                \centering
                \includegraphics[width=1.0\linewidth]{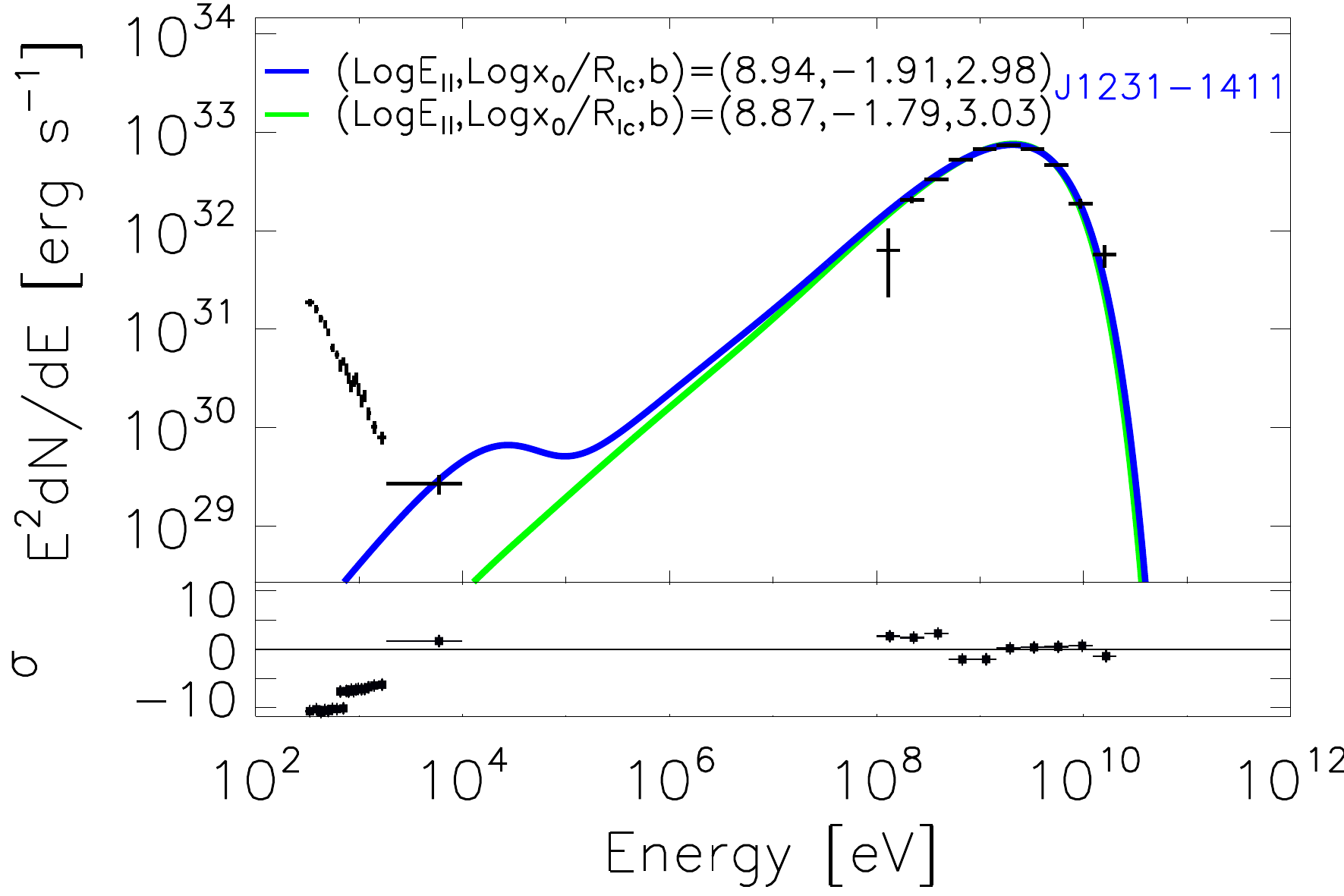}
                \label{fig:1231_xbirth_free}
            \end{subfigure}%
            \begin{subfigure}{.35\textwidth}
                \centering
                \includegraphics[width=1.0\linewidth]{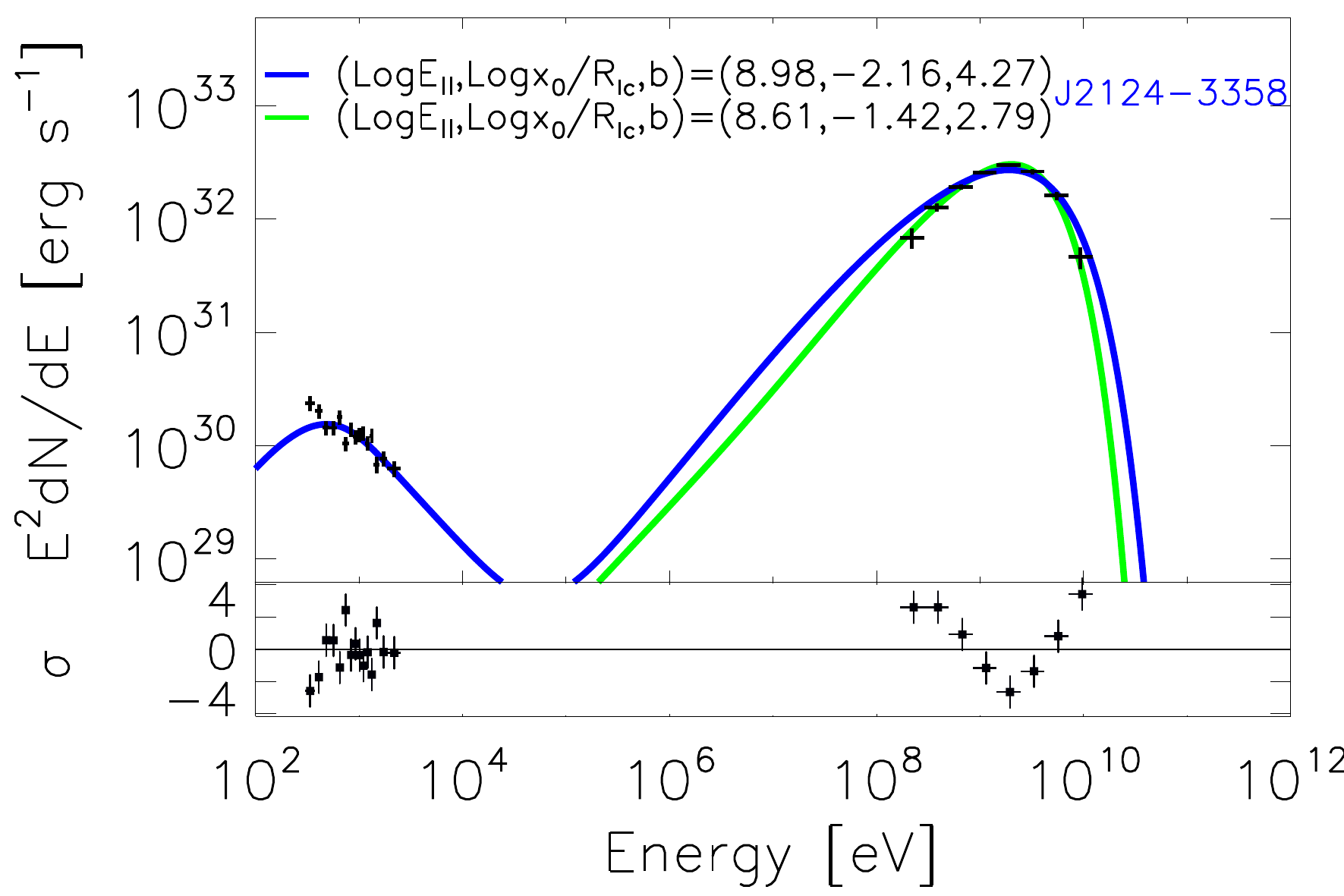}
                \label{fig:2124_xbirth_free}
            \end{subfigure}
            \caption{As in Figure \ref{fig:xbirth_scan_1} for pulsars for which
            the minimal model has $\overline{\chi^2} > 3$.
            Red lines for young pulsars PSRs and blue lines for millisecond pulsars MSPs.}
            \label{fig:xbirth_scan_2}
        \end{figure*}

Figures \ref{fig:xbirth_scan_1} and \ref{fig:xbirth_scan_2} show the fit results for the cases in which the minimal model did not show a good fit. 
The current approach is nevertheless seen to be quite good in describing the SEDs in all cases shown in Fig. \ref{fig:xbirth_scan_1}. 
Only for J2229+6114, the F-value is smaller than the critical one, meaning that in this case letting $x_{lp.in}$ free does not provide a significantly better fit than fixing it (i.e. using the minimal model).
This is reinforced by looking at the panel of Figure \ref{fig:xbirth_scan_1} corresponding to this pulsar, in which none of these spectra resembles the data points significantly better than the other.
For the other pulsars, adding a fourth parameter provides a significantly better fit.

Despite obvious improvements in some cases, the observational SEDs in Figure \ref{fig:xbirth_scan_2} are not satisfactorily reproduced by the theoretical best-fit spectra with a variable $x_{lp.in}$.
$\overline{\chi^2}$ values (shown in the lower section of Table \ref{tab:f_tests}) are relatively high and some systematic deviations are observed in the residual panels.
While in several cases (e.g. J1709-4429, J1124-5916) the new fits are better than those obtained with the minimal model; in some others (e.g. J1057-5226, J1231-1411) both fits are similar.
Most of the cases for which the minimal model required two separate accelerating
regions still require so when $x_{lp.in}$ is ranged.

We note that the values of $x_{lp.in}$ chosen by the fitting for pulsars which were already well reproduced with three parameters are very similar to those given by Eq. \eqref{eq:x_birth}.
We also note that by adding this fourth parameter, the list of pulsars that are successfully modelled by the model is enlarged (27 out the 36 pulsars studied are in this list).

The fitted values of $x_{lp.in}/R_{lc}$ show no significant correlation with $E_\parallel, x_0/R_{lc}$ or $b$, nor it shows a correlation with other parameters like the strength of the magnetic field  at the light cylinder $B_{lc}$ or the spin-down power $\dot E$.
Higher values of $x_{lp.in}/R_{lc}$, $\sim 10^{-3}$, are preferred in the few MSPs we studied in comparison with a more distributed set of values for normal pulsars, ranging from $\sim 10^{-3}$ to $\sim 10^{-7}$, weighted towards lower values.
The minimum $x_{lp.in}$ is about 70 cm for Vela and J1838, and about 1 m for MSPs, being larger for the rest. 
Physically, such small spatial scales implies that there is a small
region after a particle starts its acceleration where the 
large pitch angle particles dominate the average spectra, being contributed significantly by more evolved particles beyond.

\section{Size and location of the injection region}
\label{acc-reg}

Typical  $x_0$ values, as can be read directly from the tables, are of the order of $10^{-2.6}R_{lc}$, which for a $R_{lc}\sim 10^8$ cm is already 3 km. 
The value of our effective parameter $x_0$
represents the relevant extent within $x_{max}$ 
where the population of particles radiating towards us is mostly found once injected around $x_{in}$.
This does not mean that accelerated particles injected in $x_{in}$ do not exist beyond $x_0$, just that we do not see the radiation they emit in the framework of our model.
This does not mean either that $x_{in}$ must necessarily be unique, i.e., that there is just only one injection point where all particles are injected, and that all radiation we see comes from the associated $x_0$ scale measured from there.

\begin{figure*}
        \centering
        \includegraphics[width=0.33\textwidth]{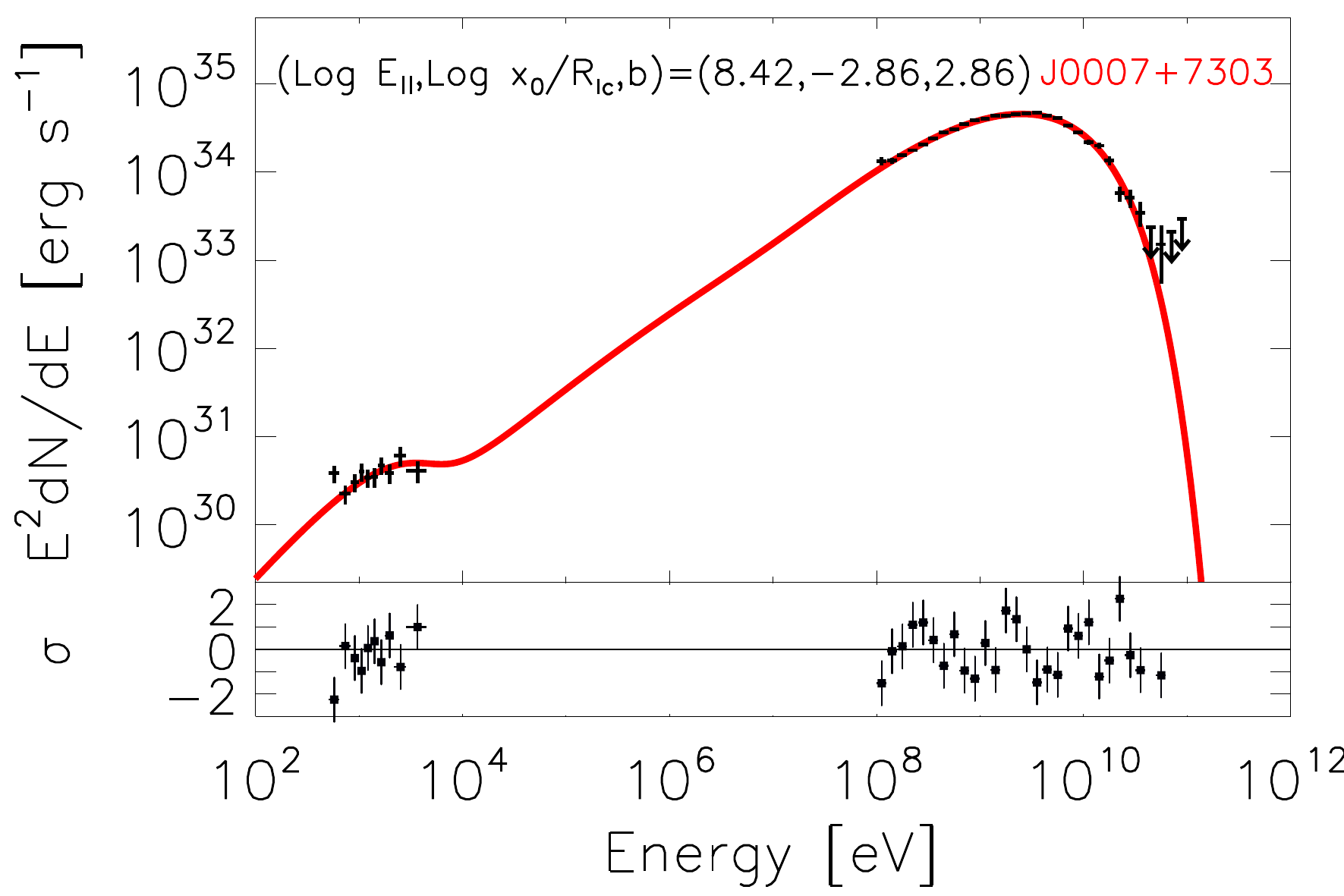}%
        \includegraphics[width=0.33\textwidth]{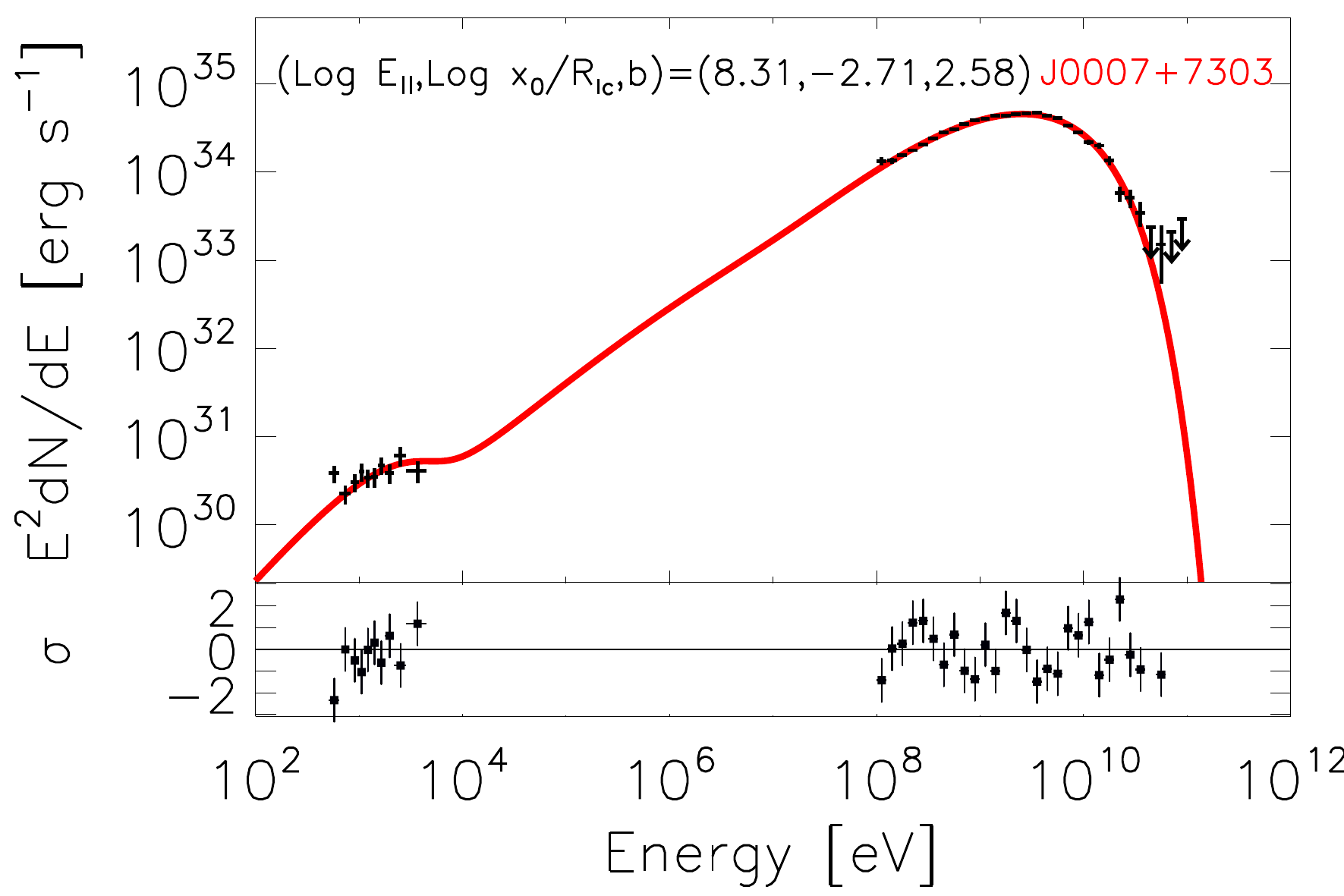}%
        \includegraphics[width=0.33\textwidth]{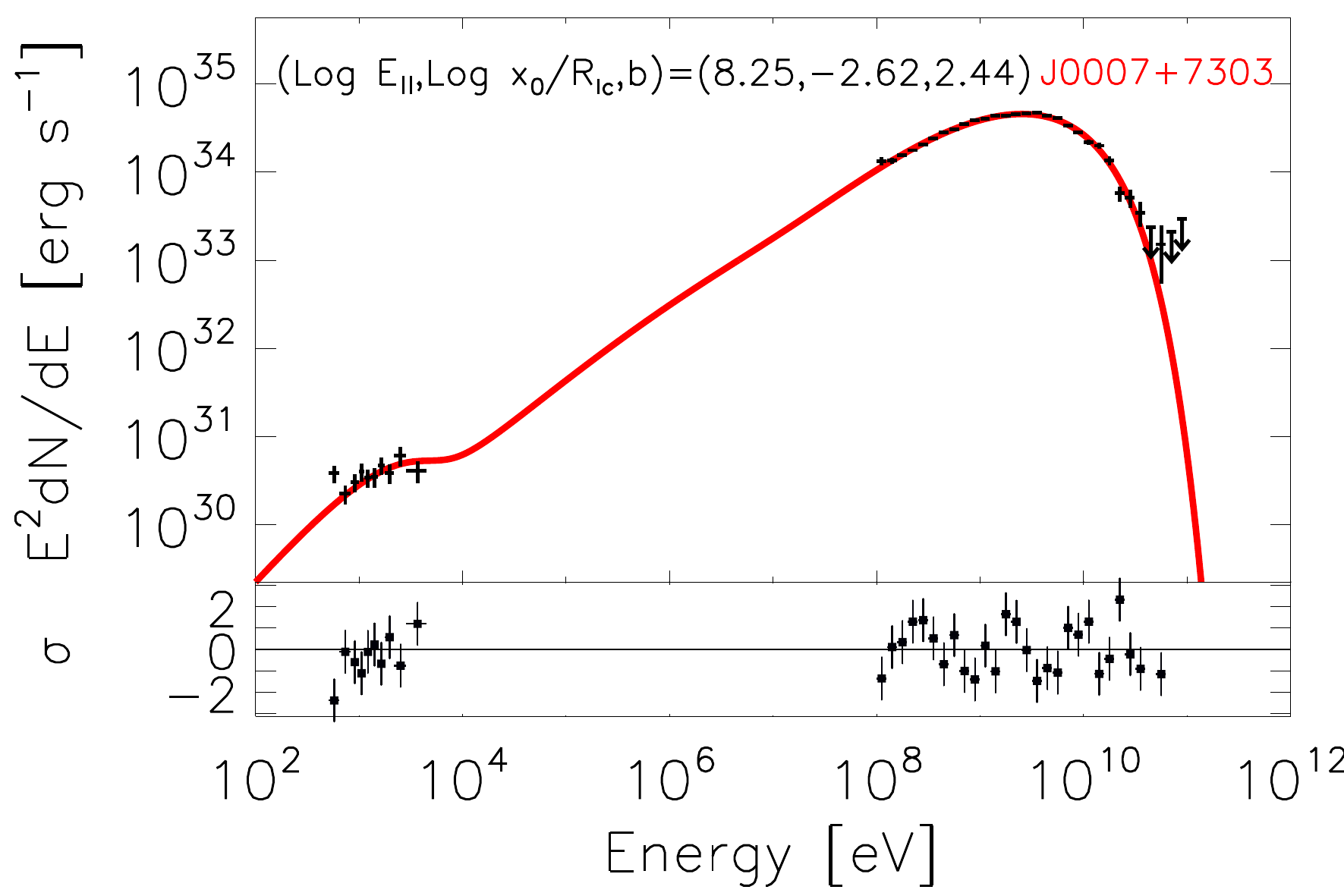}\\
        \includegraphics[width=0.33\textwidth]{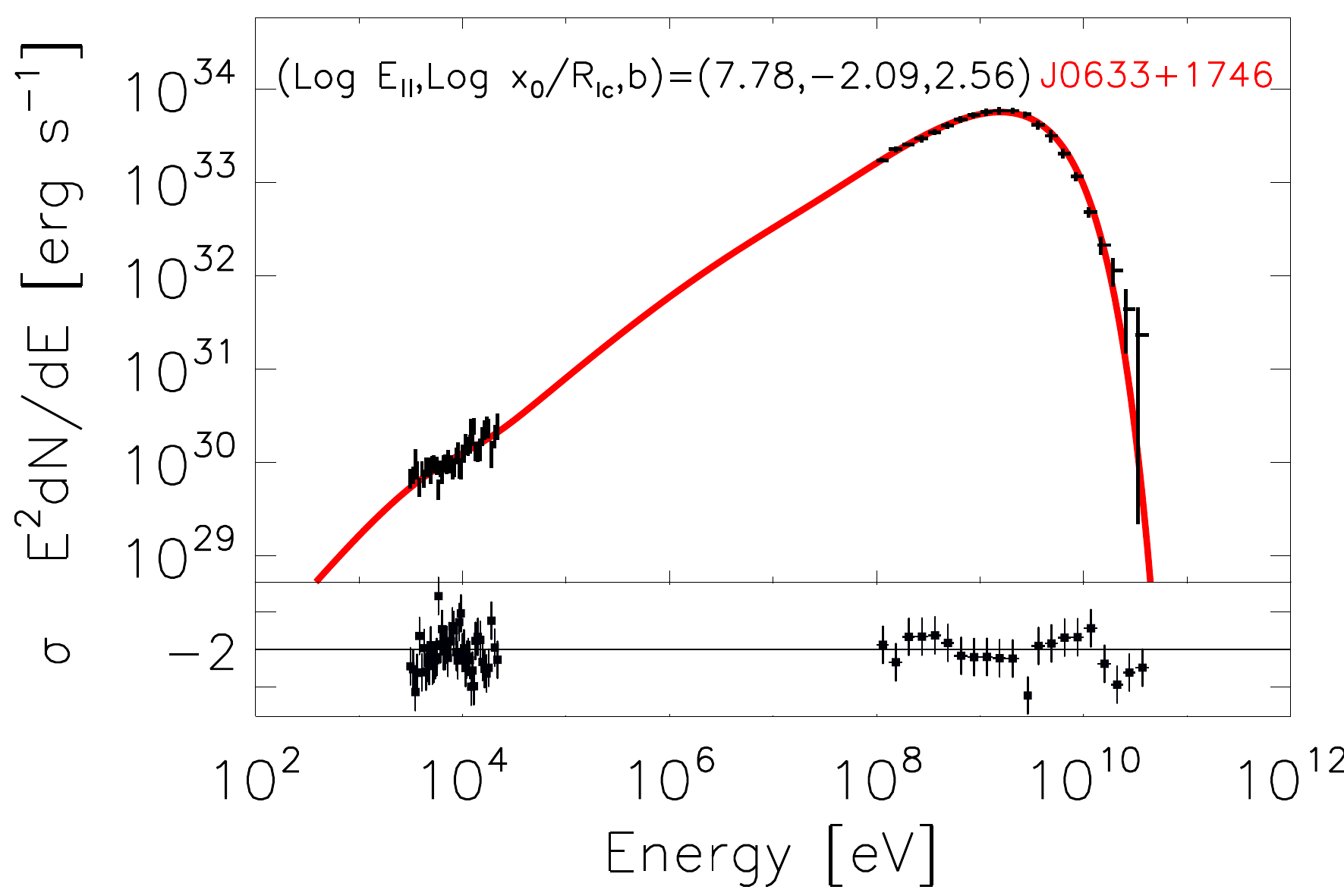}%
        \includegraphics[width=0.33\textwidth]{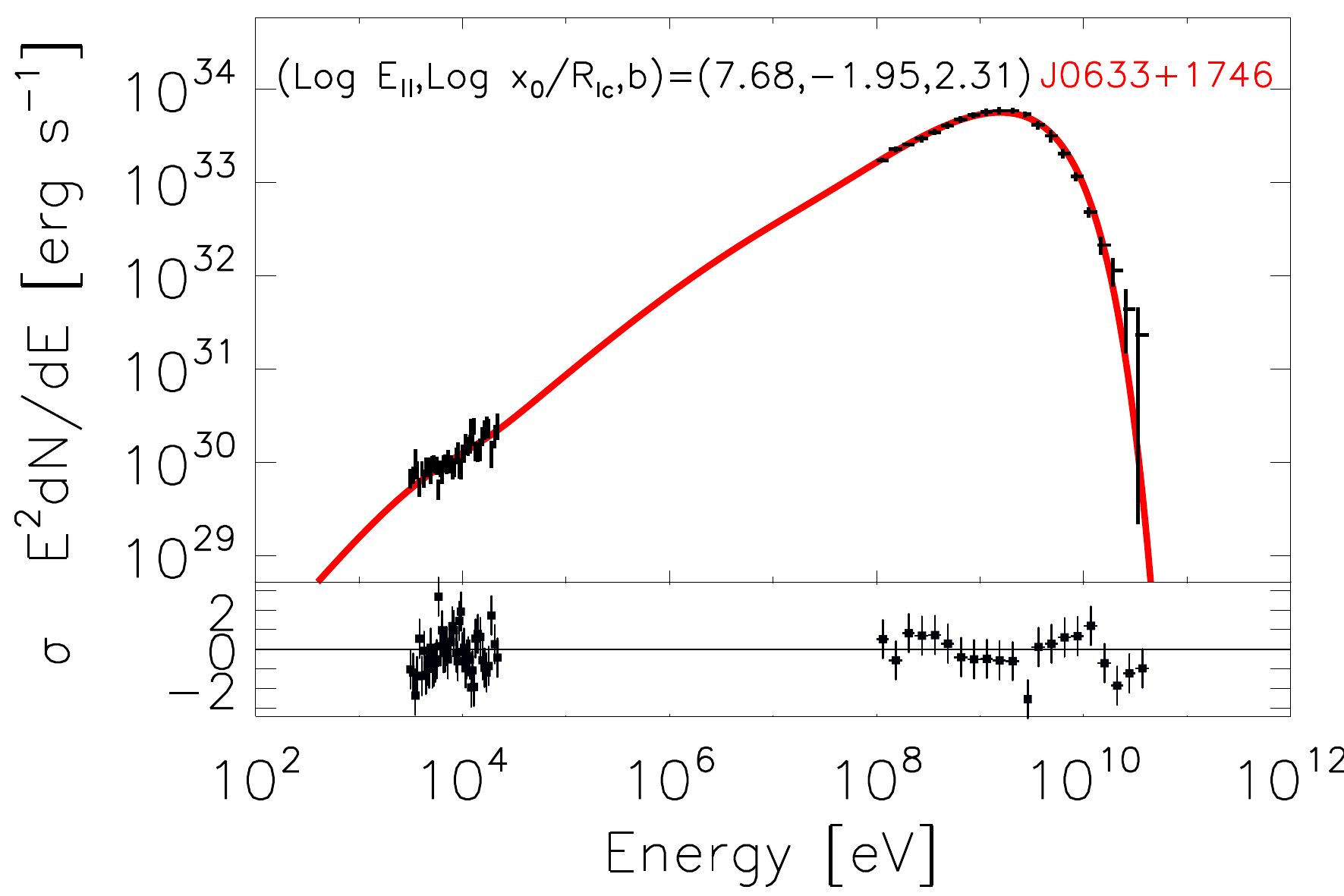}%
        \includegraphics[width=0.33\textwidth]{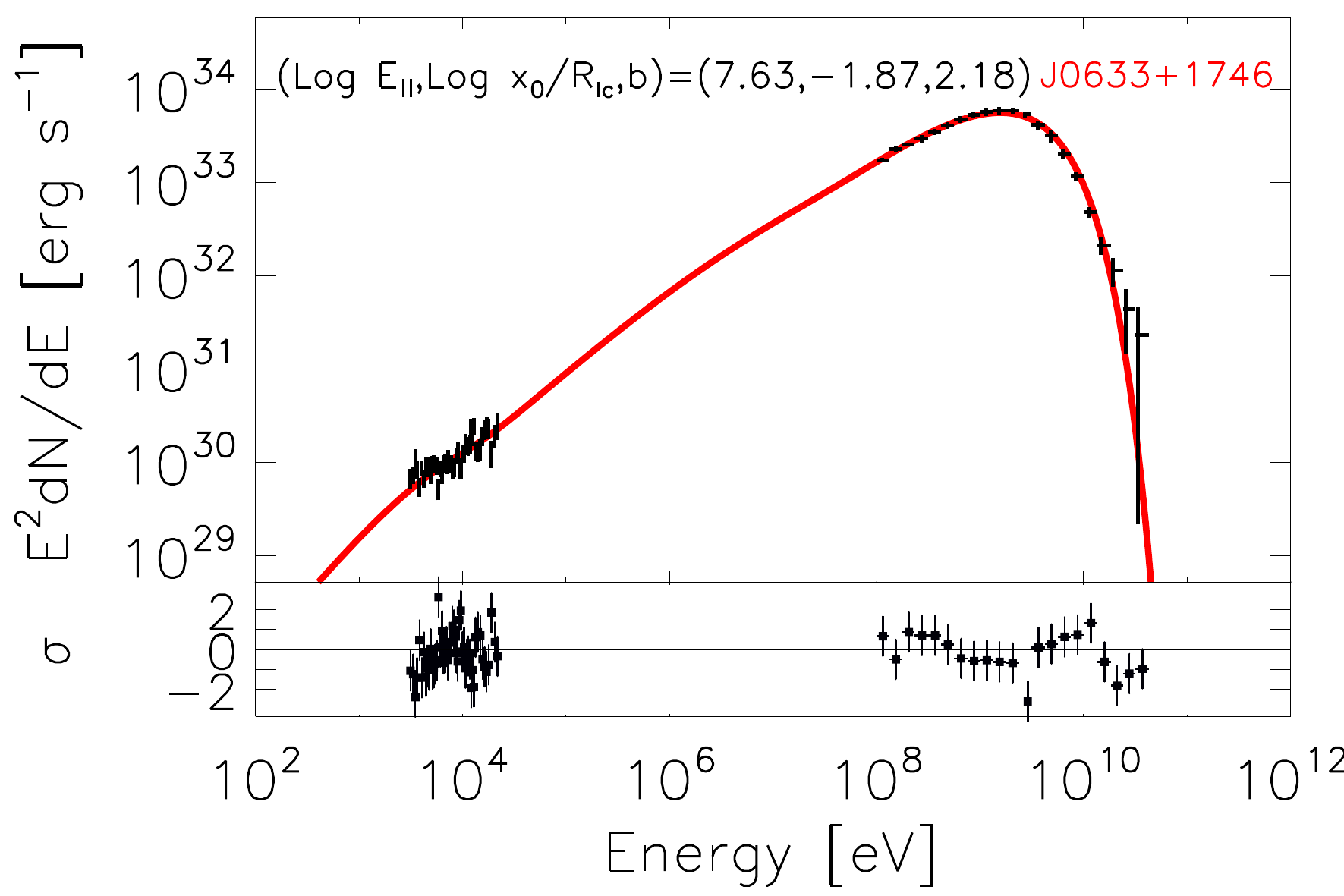}
    \caption{
    Plots of the fits of J0007+7303 and J0633+1746 (Geminga) for different values of $x_{in}$. From left to right: $x_{in} = 0.5 R_{lc}$ (as in the paper), $1.0 R_{lc}$, $1.5 R_{lc}$. 
    }
    \label{fig:systematic_study_xin}
\end{figure*}

Considering that there is only one $x_{in}$, as we have done above, $x_{in}=0.5 R_{lc}$, is a practical simplification of the model based on the fact that we have earlier proven that the results are not critically relevant for different values of $x_{in}$.
That is, if we change $x_{in}$ in a broad range around reasonable values (from say 0.2 to 1.5 $R_{lc}$), results are not significantly affected. 
As an example, Fig. \ref{fig:systematic_study_xin} proves this explicitly for some exemplary pulsars, showing how the model parameters would change by choosing a different value of $x_{in}$ where all particles are assumed to be injected.  Note the similar values of the fit parameters and the fit errors.

Thus, we can directly consider a range of injection points (many $x_{in}$, numerically) in the same region, each one treated as in the case of $x_{in}=0.5$ as we assumed before, to ultimately represent a continuous injection along $x_{max}$ (along an extent equal to 1 $R_{lc}$):
    \begin{eqnarray}
        \frac{dP_{tot}}{dE} = \int_{x_{in}}^{x_{out}} \! dx_{in}
        \int_{0}^{x_{out}} \! dx 
        \left\langle\frac{dP_{sc}}{dE}(x)\right\rangle 
        \frac{dN}{dx_{in}\,dx}(x,x_{in})
        \label{eq:fully_convolved_power_spectra_along_xmax_int}
    \end{eqnarray}

To simplify the fits, one can consider that each injection point
introduces the same number of particles (i.e., that all      $N_0(x_{in})$ are the same). Thus, in the case of continuous injection,
\begin{equation}
  \frac{dN}{dx_{in}\,dx}(x,x_{in}) = \frac{1}{(x_{out} - x_{in})}
  \frac{dN}{dx}(x,x_{in})
   \end{equation}
and $\int_{x_{in}}^{x_{out}} (dN/dx_{in} \, dx) dx=N_0(x_{in})$.

\begin{figure*}
        \centering
        \includegraphics[width=0.305\textwidth]{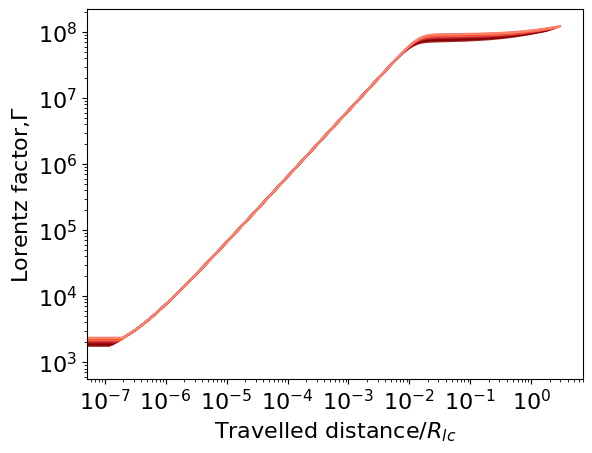}%
        \includegraphics[width=0.33\textwidth]{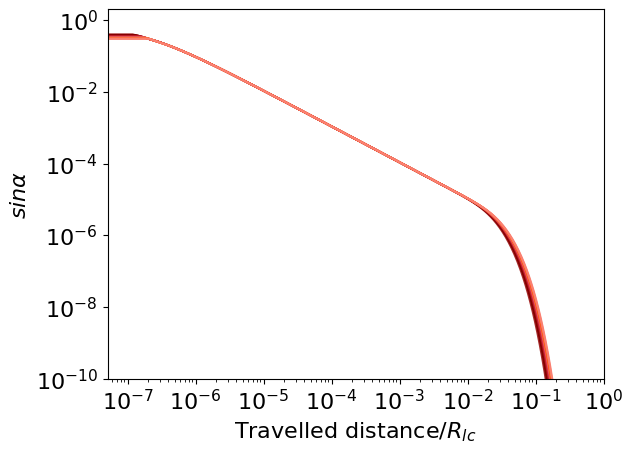}%
        \includegraphics[width=0.33\textwidth]{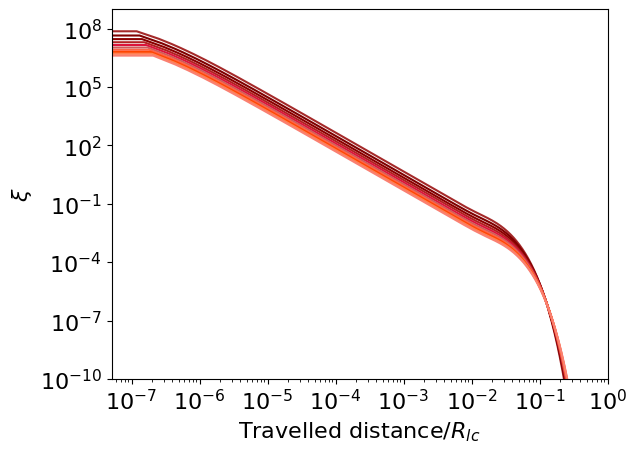}
    \caption{
    Examples of the evolution of the main properties of the particle along the trajectories for the different injection points, color coded from dark (closer to the star) to light (farther away). The x-axis represents the relative distance from the corresponding injection point, of which 10 are considered to be evenly distributed from 0.5 to 1.5 $R_{lc}$.
    }
    \label{fig:systematic_study_injection_trajectories}
\end{figure*}

Fig. \ref{fig:systematic_study_injection_trajectories}
shows how the main properties of the trajectories of particles injected along the $R_{lc}$ extent changes. In particular, 
we show the Lorentz factor $\Gamma$, the pitch angle $\alpha$,
and the synchro-curvature parameter $\xi$, for three different pulsars as an example. There is no qualitative difference in the development of these particle properties, when plotted against their relative distance to their injection points.

\begin{figure*}
        \centering
        \includegraphics[width=0.33\textwidth]{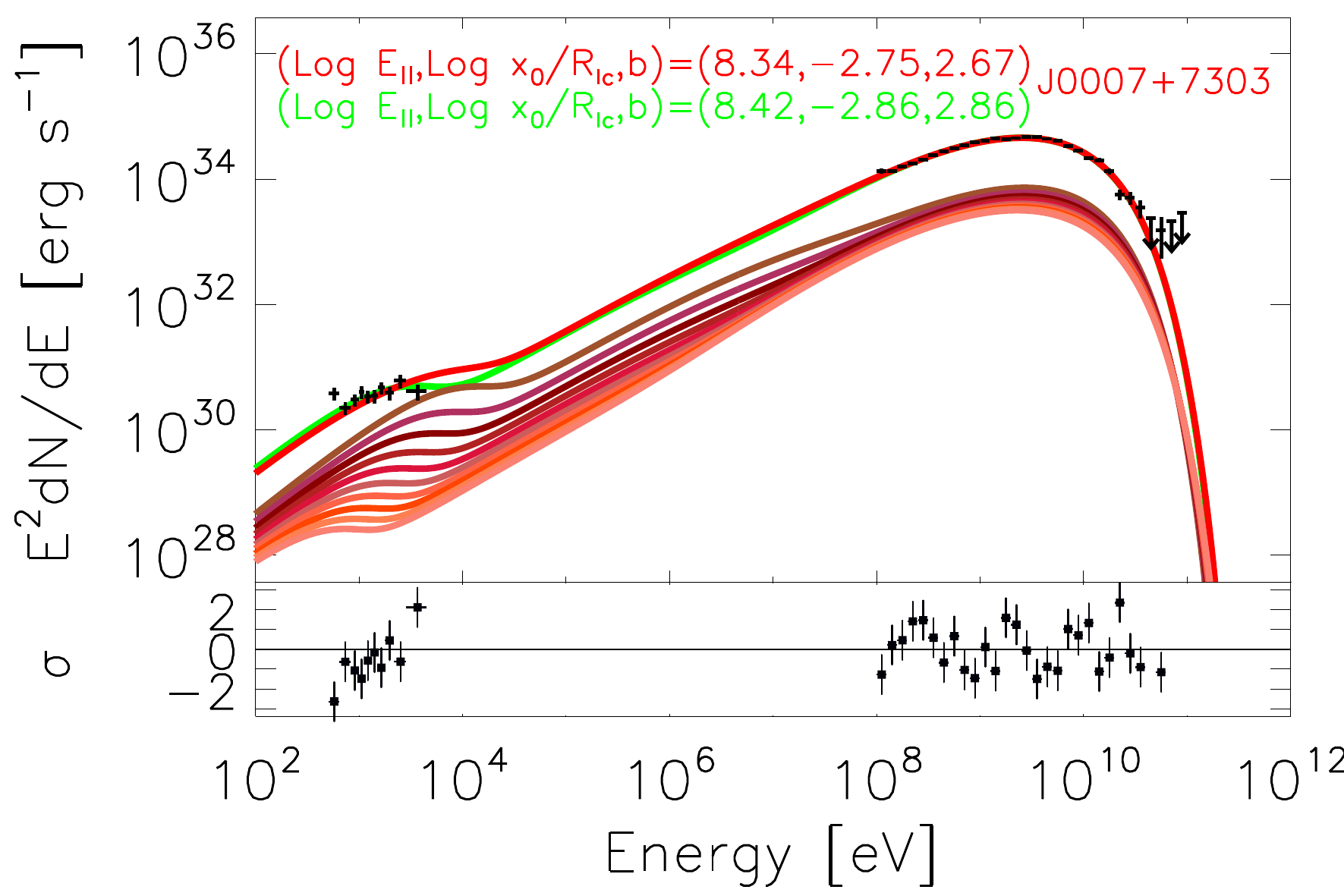}%
        \includegraphics[width=0.33\textwidth]{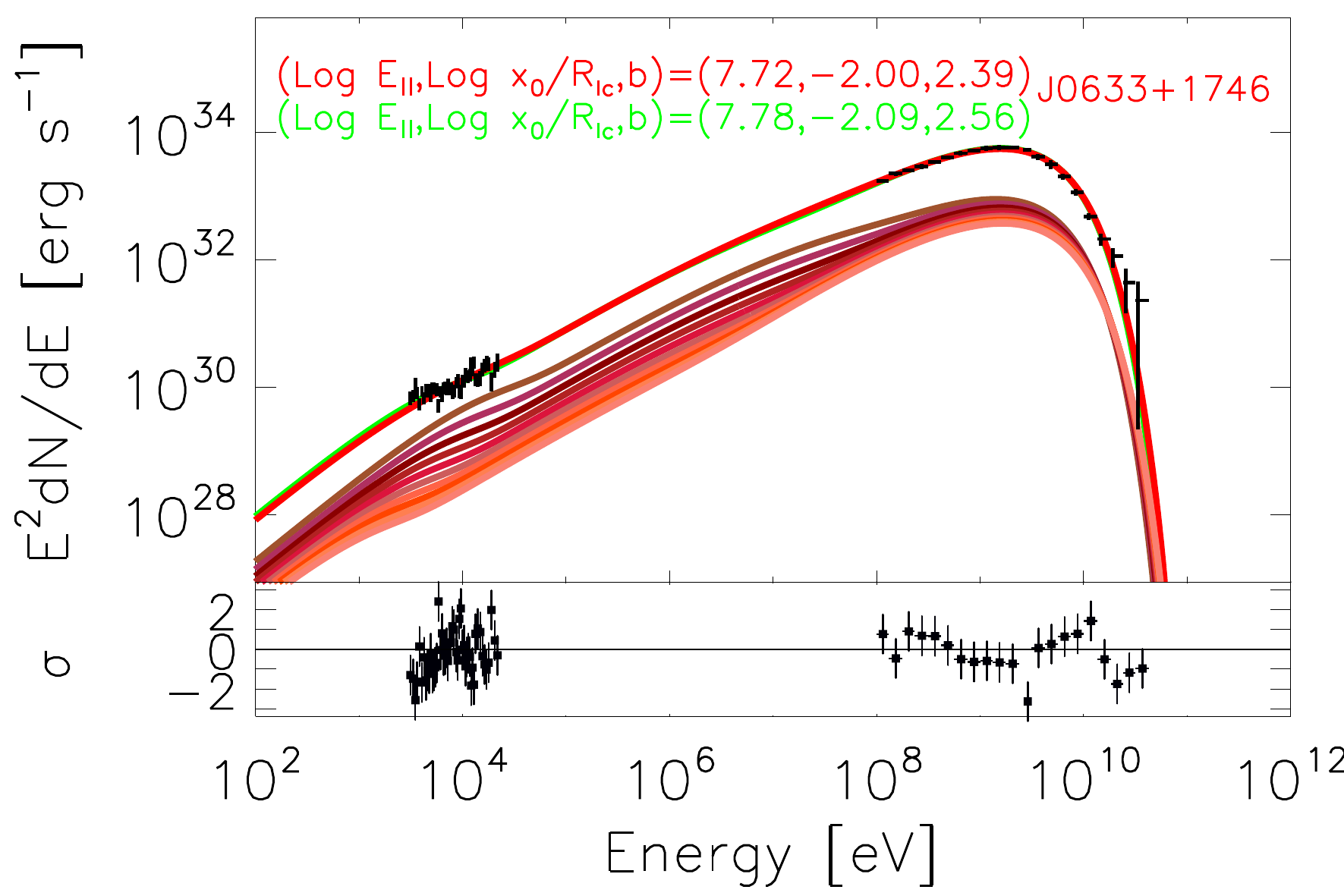}%
        \includegraphics[width=0.33\textwidth]{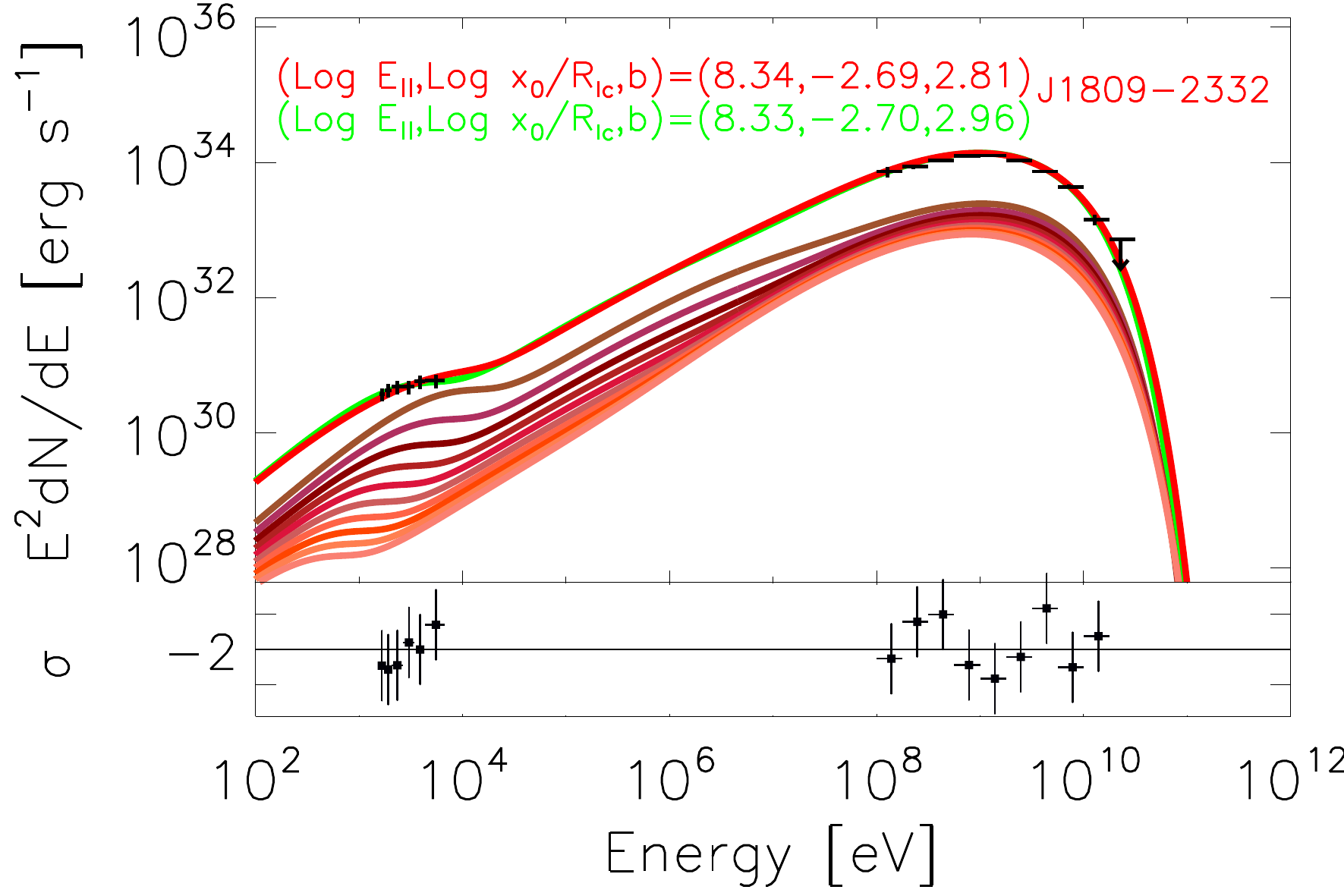}
    \caption{Examples of the SEDs comparing the results obtained with a single injection point $x_{in}=0.5 R_{lc}$ (green lines), and with 10 injection points evenly distributed from 0.5 to 1.5 $R_{lc}$ (red lines).
    We also show the contributions of each of the injection points
    (color coded from dark (closer to the star) to light (farther away). 
    Residuals are those of the SED obtained with 10 injection points evenly distributed from 0.5 to 1.5 $R_{lc}$.
    }
    \label{fig:systematic_study_injection_SED}
\end{figure*}

\begin{figure*}
        \centering
        \includegraphics[width=0.33\textwidth]{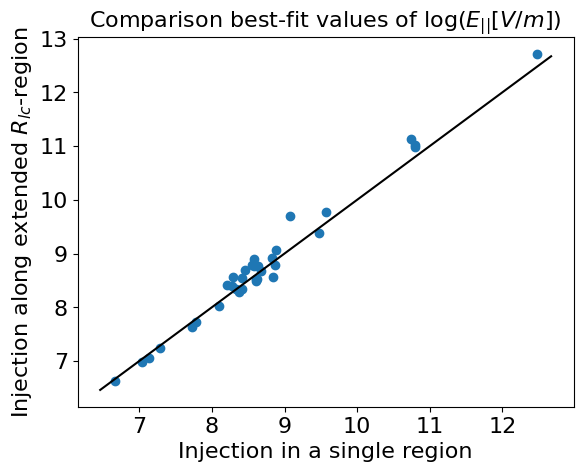}%
        \includegraphics[width=0.33\textwidth]{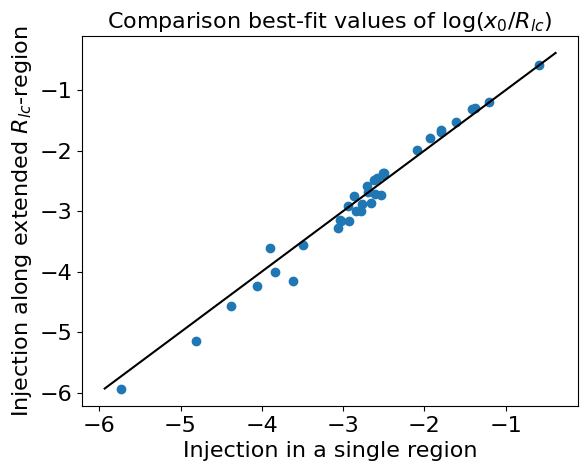}\\
        \includegraphics[width=0.33\textwidth]{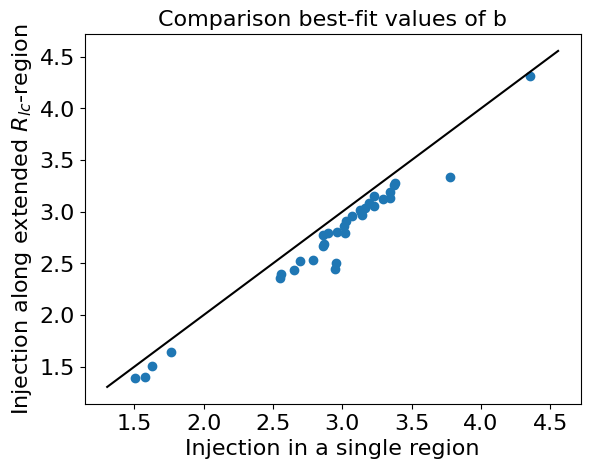}%
        \includegraphics[width=0.33\textwidth]{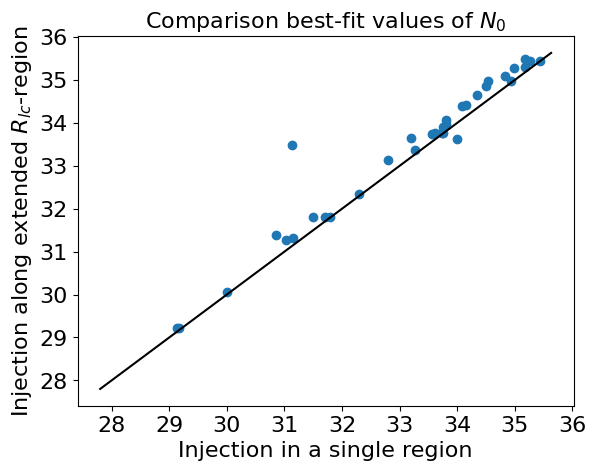}
    \caption{Comparison between best-fitting parameters 
    using a single injection point and those obtained using a distributed injection along a region of size $R_{lc}$ for all pulsars in our sample. Black lines represent equal values. The difference in the value of $b$ is 
    and discussed in the text.}
    \label{fig:comparing_injection}
\end{figure*}

Fig. \ref{fig:systematic_study_injection_SED} shows that, when considering several injection points instead of only 1, the values of $E_{||}$ and $x_0$  are essentially unchanged, whereas the value of $b$ gets an averaged value between that resulting in the case when all particles are injected at the first (last) point only.
This happens to compensate the distance effect reduction of the magnetic field at injection, so that the local magnetic field is similar in all cases. The magnetic gradient $b$ is smaller, i.e., $B$ is larger, the closer to the star are the injection points. This mimics what happens with millisecond pulsars, for which typically larger magnetic gradients are obtained to accommodate similar local magnetic fields in the accelerating region.
Fig. \ref{fig:comparing_injection} finally provides a visual summary, showing 
how the fitting parameters obtained using a distributed injection along a region of size $R_{lc}$ compare for all the pulsars in the sample.

Particles can thus be assumed to be injected along the whole extent of the acceleration region, of size $R_{lc}$ or larger, and the fit results are essentially unchanged. They tell us, as before, that the
values needed for $x_0$ are small in comparison to $R_{lc}$.

Note that a direct comparison between PIC simulations, e.g. \cite{Phillippov14, Chen14, Cerutti19}, and our effective model is tricky. Both have significant caveats and different defining points. 
The connection between our effective spectral model --and the value of the fitted parameters-- and the light curves is currently under study. Our spectral model does not contain any geometry (it is 1D). 
In order to do this, we have already introduced the methodology needed in the case of no  radiation (see \cite{light_curves}) and are currently adapting the method to encompass synchro-curvature emission. We expect to report on this in the future.

\section{Concluding remarks}
\label{conclusions}

We have improved the physical and numerical treatment of the injection and the emission of the particles traveling across the accelerating region in a synchro-curvature radiation model introduced before.
We have incorporated into the model a better description of the innermost region of particle's trajectories by averaging over the emitted spectra of particles injected at different points.
The new approach allowed to describe a few pulsar SEDs (e.g., J0357+3205 and J1826-1256) where previously was not possible. 
However, in general, while obtaining an improvement on the fits of most pulsars, overall results remained similar.
The model is able to describe reasonably well most of the pulsars, across 8 orders of magnitude from X-rays to gamma-rays, with just three parameters related to the local value of magnetic field in the accelerating region --represented by the magnetic gradient, the parallel electric field, and a lengthscale of the emitting region. 
As before, the minority of pulsars for which the model  did not provide a good fit, can be successfully described with a simple extension of the model in which two acceleration regions differing in their geometrical properties are considered.
We have also here explicitly shown how, whereas the accelerating region can be large in extent, the observationally-relevant spatial scales for particles emitting towards us once injected at a given point is small in comparison.

At a numerical level, we have implemented a significant change in the spectral fitting procedure by using as minimization algorithm the Nelder-Mead method (instead of the earlier search by brute-force gridding). 
The new fitting procedure has decreased the computational time required for the systematic study of the whole population of high-energy pulsars by more than one order of magnitude.  
This improvement opens the gate to a range of new studies that were not possible before due to computational limitations.
Among these, what we presented here already as well, it allows an exploration over the spatial extent of the 
particle's trajectories injected at a given point where the fresh particle spectra (with the largest pitch angles) dominates.
We did this by assuming it as an extra parameter of the model, $x_{lp.in}$, instead of taking it as fixed estimate.
Setting the parameter $x_{lp.in}$ as the fourth free model parameter --instead of estimating it from the trajectory of particles- allows to fit well a few more pulsars. 
However, we find that for the majority of our sample, an effective value $x_{lp.in}(P)$ as we assumed before already gives very good fits (allowing a more agile and fast fitting), 
and that the estimate is reasonably good.

We have thus confirmed that the synchro-curvature description of the pulsar's SED
requires small spatial scales. On the one hand, $x_{lp.in}\sim 10^{-4}-10^{-6}$ $R_{lc}$, where the spectrum emitted by particles injected at a given point is dominated by the large pitch angle emission via synchrotron. On the other hand,  
$x_0 \sim 10^{-2}-10^{-3}$ $R_{lc}$, 
which represent 
that after injection particles quickly stops being observationally-relevant for the observer.
These scales should not be confused to the extent of the acceleration region, and do not imply that particle 
acceleration or injection occur only on them. We have also explicitly shown that, if we consider particle injection occurring all along the acceleration region (i.e., a range of $x_{in}$ along $x_{max}$, herein assumed to be at least $R_{lc}$), the fits are practically unchanged.

In closing, we note the small value of $E_{\parallel}$ singled out for J0357+3205 and J2055+2539 in Table \ref{tab:best_fit_parameters}.
Smaller electric fields will accelerate particles to lower values of $\Gamma$, and the SED can peak at MeV energies, still with non-negligible flux at GeV range.
Very few of these MeV-pulsars are currently known. 
One example is J1846-0258, which has a small best-fit value of $E_{\parallel}$ and whose SED peaks at $\sim 1$ MeV \citep{Kuiper2018}.
We argue that J0357+3205 and J2055+2539 are  MeV-pulsar candidates and are suggested for exhaustive observations in this energy band.

\section*{Acknowledgements}
This work has been supported by the grants
PID2021-124581OB-I00, PGC2018-095512-B-I00, as well as 
the Spanish program Unidad de Excelencia ``María de
Maeztu'' CEX2020-001058-M. 
DIP has been supported by CSIC program JAE-ICU.
DV is funded by the European Research Council (ERC) under the European Union’s Horizon 2020 research and innovation programme (ERC Starting Grant IMAGINE, No. 948582). 
DFT also acknowledges USTC and  the Chinese Academy of Sciences Presidential Fellowship Initiative 2021VMA0001.

\section*{Data Availability}

No new observational data is herein presented. The dataset used is publicly available already via \cite{CotiZelati20}.
Any additional theoretical detail required is available from the authors on reasonable request.

\bibliography{sc_emitting_regions}

\begin{thebibliography}{}
\makeatletter
\relax
\def\mn@urlcharsother{\let\do\@makeother \do\$\do\&\do\#\do\^\do\_\do\%\do\~}
\def\mn@doi{\begingroup\mn@urlcharsother \@ifnextchar [ {\mn@doi@}
  {\mn@doi@[]}}
\def\mn@doi@[#1]#2{\def\@tempa{#1}\ifx\@tempa\@empty \href
  {http://dx.doi.org/#2} {doi:#2}\else \href {http://dx.doi.org/#2} {#1}\fi
  \endgroup}
\def\mn@eprint#1#2{\mn@eprint@#1:#2::\@nil}
\def\mn@eprint@arXiv#1{\href {http://arxiv.org/abs/#1} {{\tt arXiv:#1}}}
\def\mn@eprint@dblp#1{\href {http://dblp.uni-trier.de/rec/bibtex/#1.xml}
  {dblp:#1}}
\def\mn@eprint@#1:#2:#3:#4\@nil{\def\@tempa {#1}\def\@tempb {#2}\def\@tempc
  {#3}\ifx \@tempc \@empty \let \@tempc \@tempb \let \@tempb \@tempa \fi \ifx
  \@tempb \@empty \def\@tempb {arXiv}\fi \@ifundefined
  {mn@eprint@\@tempb}{\@tempb:\@tempc}{\expandafter \expandafter \csname
  mn@eprint@\@tempb\endcsname \expandafter{\@tempc}}}

\bibitem[\protect\citeauthoryear{Abdo et~al.,}{Abdo et~al.}{2013}]{2fpc}
Abdo A.~A.,  et~al., 2013, \mn@doi [The Astrophysical Journal Supplement
  Series] {10.1088/0067-0049/208/2/17}, 208, 17

\bibitem[\protect\citeauthoryear{{Acciari} et~al.,}{{Acciari}
  et~al.}{2021}]{0218_analysis}
{Acciari} V.~A.,  et~al., 2021, \mn@doi [\apj] {10.3847/1538-4357/ac20d7},
  \href {https://ui.adsabs.harvard.edu/abs/2021ApJ...922..251A} {922, 251}

\bibitem[\protect\citeauthoryear{{Arons}}{{Arons}}{1983}]{arons83}
{Arons} J.,  1983, \mn@doi [\apj] {10.1086/160771}, \href
  {https://ui.adsabs.harvard.edu/abs/1983ApJ...266..215A} {266, 215}

\bibitem[\protect\citeauthoryear{{Avni}}{{Avni}}{1976}]{avni_factor}
{Avni} Y.,  1976, \mn@doi [\apj] {10.1086/154870}, \href
  {https://ui.adsabs.harvard.edu/abs/1976ApJ...210..642A} {210, 642}

\bibitem[\protect\citeauthoryear{{Cerutti}}{{Cerutti}}{2019}]{Cerutti19}
{Cerutti} B.,  2019, \mn@doi [Rendiconti Lincei. Scienze Fisiche e Naturali]
  {10.1007/s12210-019-00864-y}, \href
  {https://ui.adsabs.harvard.edu/abs/2019RLSFN..30S..89C} {30, 89}

\bibitem[\protect\citeauthoryear{{Chen} \& {Beloborodov}}{{Chen} \&
  {Beloborodov}}{2014}]{Chen14}
{Chen} A.~Y.,  {Beloborodov} A.~M.,  2014, \mn@doi [\apjl]
  {10.1088/2041-8205/795/1/L22}, \href
  {https://ui.adsabs.harvard.edu/abs/2014ApJ...795L..22C} {795, L22}

\bibitem[\protect\citeauthoryear{{Cheng} \& {Zhang}}{{Cheng} \&
  {Zhang}}{1996}]{cheng_zhang}
{Cheng} K.~S.,  {Zhang} J.~L.,  1996, \mn@doi [\apj] {10.1086/177239}, \href
  {https://ui.adsabs.harvard.edu/abs/1996ApJ...463..271C} {463, 271}

\bibitem[\protect\citeauthoryear{{Cheng}, {Ho}  \& {Ruderman}}{{Cheng}
  et~al.}{1986}]{Cheng86}
{Cheng} K.~S.,  {Ho} C.,   {Ruderman} M.,  1986, \mn@doi [\apj]
  {10.1086/163829}, \href
  {https://ui.adsabs.harvard.edu/abs/1986ApJ...300..500C} {300, 500}

\bibitem[\protect\citeauthoryear{{Coti Zelati}, {Torres}, {Li}  \&
  {Vigan{\`o}}}{{Coti Zelati} et~al.}{2020}]{CotiZelati20}
{Coti Zelati} F.,  {Torres} D.~F.,  {Li} J.,   {Vigan{\`o}} D.,  2020, \mn@doi
  [\mnras] {10.1093/mnras/stz3485}, \href
  {https://ui.adsabs.harvard.edu/abs/2020MNRAS.492.1025C} {492, 1025}

\bibitem[\protect\citeauthoryear{{Goldreich} \& {Julian}}{{Goldreich} \&
  {Julian}}{1969}]{pulsar_electrodynamics}
{Goldreich} P.,  {Julian} W.~H.,  1969, \mn@doi [\apj] {10.1086/150119}, \href
  {https://ui.adsabs.harvard.edu/abs/1969ApJ...157..869G} {157, 869}

\bibitem[\protect\citeauthoryear{{H.~E.~S.~S. Collaboration}
  et~al.,}{{H.~E.~S.~S. Collaboration} et~al.}{2018}]{HESSVela200}
{H.~E.~S.~S. Collaboration} et~al., 2018, \mn@doi [\aap]
  {10.1051/0004-6361/201732153}, \href
  {https://ui.adsabs.harvard.edu/abs/2018A&A...620A..66H} {620, A66}

\bibitem[\protect\citeauthoryear{{Harding}, {Stern}, {Dyks}  \&
  {Frackowiak}}{{Harding} et~al.}{2008}]{Harding2008}
{Harding} A.~K.,  {Stern} J.~V.,  {Dyks} J.,   {Frackowiak} M.,  2008, \mn@doi
  [\apj] {10.1086/588037}, \href
  {https://ui.adsabs.harvard.edu/abs/2008ApJ...680.1378H} {680, 1378}

\bibitem[\protect\citeauthoryear{{Kuiper}, {Hermsen}  \& {Dekker}}{{Kuiper}
  et~al.}{2018}]{Kuiper2018}
{Kuiper} L.,  {Hermsen} W.,   {Dekker} A.,  2018, \mn@doi [\mnras]
  {10.1093/mnras/stx3128}, \href
  {https://ui.adsabs.harvard.edu/abs/2018MNRAS.475.1238K} {475, 1238}

\bibitem[\protect\citeauthoryear{{Li}, {Torres}, {Coti Zelati}, {Papitto},
  {Kerr}  \& {Rea}}{{Li} et~al.}{2018}]{Li18}
{Li} J.,  {Torres} D.~F.,  {Coti Zelati} F.,  {Papitto} A.,  {Kerr} M.,   {Rea}
  N.,  2018, \mn@doi [\apjl] {10.3847/2041-8213/aae92b}, \href
  {https://ui.adsabs.harvard.edu/abs/2018ApJ...868L..29L} {868, L29}

\bibitem[\protect\citeauthoryear{{Lyubarskii} \& {Petrova}}{{Lyubarskii} \&
  {Petrova}}{1998}]{Lyubarskii1998}
{Lyubarskii} Y.~E.,  {Petrova} S.~A.,  1998, \aap, \href
  {https://ui.adsabs.harvard.edu/abs/1998A&A...337..433L} {337, 433}

\bibitem[\protect\citeauthoryear{{MAGIC Collaboration} et~al.,}{{MAGIC
  Collaboration} et~al.}{2020}]{MAGIC-2020-Geminga}
{MAGIC Collaboration} et~al., 2020, \mn@doi [\aap]
  {10.1051/0004-6361/202039131}, \href
  {https://ui.adsabs.harvard.edu/abs/2020A&A...643L..14M} {643, L14}

\bibitem[\protect\citeauthoryear{{Manchester}, {Hobbs}, {Teoh}  \&
  {Hobbs}}{{Manchester} et~al.}{2005}]{ATNF-Catalog}
{Manchester} R.~N.,  {Hobbs} G.~B.,  {Teoh} A.,   {Hobbs} M.,  2005, \mn@doi
  [\aj] {10.1086/428488}, \href
  {https://ui.adsabs.harvard.edu/abs/2005AJ....129.1993M} {129, 1993}

\bibitem[\protect\citeauthoryear{Nelder \& Mead}{Nelder \&
  Mead}{1965}]{nelder_mead}
Nelder J.~A.,  Mead R.,  1965, \mn@doi [Comput. J.] {10.1093/comjnl/7.4.308},
  7, 308

\bibitem[\protect\citeauthoryear{{Philippov} \& {Spitkovsky}}{{Philippov} \&
  {Spitkovsky}}{2014}]{Phillippov14}
{Philippov} A.~A.,  {Spitkovsky} A.,  2014, \mn@doi [\apjl]
  {10.1088/2041-8205/785/2/L33}, \href
  {https://ui.adsabs.harvard.edu/abs/2014ApJ...785L..33P} {785, L33}

\bibitem[\protect\citeauthoryear{Press, Teukolsky, Vetterling  \&
  Flannery}{Press et~al.}{1992}]{numerical_recipes}
Press W.~H.,  Teukolsky S.~A.,  Vetterling W.~T.,   Flannery B.~P.,  1992,
  Numerical Recipes in C (2nd Ed.): The Art of Scientific Computing.
Cambridge University Press, USA

\bibitem[\protect\citeauthoryear{{Torres}}{{Torres}}{2018}]{diego_solo}
{Torres} D.~F.,  2018, \mn@doi [Nature Astronomy] {10.1038/s41550-018-0384-5},
  \href {https://ui.adsabs.harvard.edu/abs/2018NatAs...2..247T} {2, 247}

\bibitem[\protect\citeauthoryear{{Torres}, {Vigan{\`o}}, {Coti Zelati}  \&
  {Li}}{{Torres} et~al.}{2019}]{systematic_2019}
{Torres} D.~F.,  {Vigan{\`o}} D.,  {Coti Zelati} F.,   {Li} J.,  2019, \mn@doi
  [\mnras] {10.1093/mnras/stz2403}, \href
  {https://ui.adsabs.harvard.edu/abs/2019MNRAS.489.5494T} {489, 5494}

\bibitem[\protect\citeauthoryear{{Uzdensky}}{{Uzdensky}}{2003}]{Uzdensky2003}
{Uzdensky} D.~A.,  2003, \mn@doi [\apj] {10.1086/378849}, \href
  {https://ui.adsabs.harvard.edu/abs/2003ApJ...598..446U} {598, 446}

\bibitem[\protect\citeauthoryear{{Vigan{\`o}} \& {Torres}}{{Vigan{\`o}} \&
  {Torres}}{2019}]{light_curves}
{Vigan{\`o}} D.,  {Torres} D.~F.,  2019, \mn@doi [\mnras]
  {10.1093/mnras/stz2685}, \href
  {https://ui.adsabs.harvard.edu/abs/2019MNRAS.490.1437V} {490, 1437}

\bibitem[\protect\citeauthoryear{Viganò \& Torres}{Viganò \&
  Torres}{2015}]{Vigan_2015}
Viganò D.,  Torres D.~F.,  2015, \mn@doi [\mnras] {10.1093/mnras/stv579}, 449,
  3755–3765

\bibitem[\protect\citeauthoryear{Viganò, Torres, Hirotani  \& Pessah}{Viganò
  et~al.}{2015a}]{compact_formulae}
Viganò D.,  Torres D.~F.,  Hirotani K.,   Pessah M.~E.,  2015a, \mn@doi
  [\mnras] {10.1093/mnras/stu2456}, 447, 1164–1172

\bibitem[\protect\citeauthoryear{Viganò, Torres, Hirotani  \& Pessah}{Viganò
  et~al.}{2015b}]{outer_gap_model_paper_1}
Viganò D.,  Torres D.~F.,  Hirotani K.,   Pessah M.~E.,  2015b, \mn@doi
  [\mnras] {10.1093/mnras/stu2564}, 447, 2631–2648

\bibitem[\protect\citeauthoryear{Viganò, Torres, Hirotani  \& Pessah}{Viganò
  et~al.}{2015c}]{outer_gap_model_paper_2}
Viganò D.,  Torres D.~F.,  Hirotani K.,   Pessah M.~E.,  2015c, \mn@doi
  [\mnras] {10.1093/mnras/stu2565}, 447, 2649–2657

\bibitem[\protect\citeauthoryear{Viganò, Torres  \& Martín}{Viganò
  et~al.}{2015d}]{Vigan_2015b}
Viganò D.,  Torres D.~F.,   Martín J.,  2015d, \mn@doi [\mnras]
  {10.1093/mnras/stv1582}, 453, 2600–2622

\makeatother
\end{thebibliography}
\bibliographystyle{mnras}

\appendix
\section{Reminder of basic formulae for the synchro-curvature spectrum}\label{app:model}

The radiation emitted by one charged particle at each position of the trajectory, following the complete synchro-curvature formulae, \cite{cheng_zhang, compact_formulae}:
\begin{ceqn}
\begin{equation}
    \frac{d P_{sc}}{dE} = \frac{\sqrt{3}(Ze)^2\Gamma y}{4 \pi \hbar r_{eff}} \left[(1+z)F(y) - (1-z)K_{2/3}(y)\right]
    \label{eq:sc_energy_spectra}
\end{equation}
\end{ceqn}
with the several quantities defined as:
\begin{ceqn}
    \begin{align}
        F(y) &= \int^{\infty}_y K_{5/3}(y')dy' \\
        r_{gyr} &= \frac{m c^2 \Gamma \sin{\alpha}}{e B} \label{eq:f_y}\\
        \xi &= \frac{r_c}{r_{gyr}} \frac{\sin^2{\alpha}}{\cos^2{\alpha}} \label{eq:xi_def} \\\
        r_{eff} &= \frac{r_c}{\cos^2{\alpha}} \left(1 + \xi + \frac{r_{gyr}}{r_c} \right)^{-1} \label{eq:effective_radius} \\
        Q_2^2 &= \frac{\cos^4{\alpha}}{r_c^2} \left[ 1 + 3\xi + \xi^2 +\frac{r_{gyr}}{r_c} \right] \label{eq:q2}\\
        E_c &= \frac{3}{2} \hbar c Q_2 \Gamma^3 \label{eq:e_c} \\
        z &= (Q_2r_{eff})^{-2} \label{eq:z}\\
        y &= \frac{E}{E_c}
        \label{eq:y}
    \end{align}
\end{ceqn}
where $K_n$ are the modified Bessel functions of the second kind of index $n$, the solutions of the Bessel equation with complex argument.
$r_{gyr}$ is the Larmor radius, $e$ and $m$ are the charge and rest mass of a lepton, $c$ is the speed of light, $\alpha$ is the pitch angle and $E$ and $E_c$ are the photon energy and the characteristic energy of the emitted radiation, respectively.
$\xi$ is the synchro-curvature parameter, which indicates whether the emission is dominated by synchrotron or by curvature radiation, or if it is a mixture of both.

\bsp	
\label{lastpage}
\end{document}